\documentclass[a4paper,reprint,twocolumn,superscriptaddress,secnumarabic,amsmath,amssymb,aps]{revtex4-1}

\usepackage{amssymb}
\usepackage{amsmath}
\usepackage{braket}
\usepackage{color}
\usepackage{balance}
\usepackage{graphicx}

\begin{document}

\title{Mapping confinement potentials and charge densities of interacting quantum systems using conditional generative adversarial networks}

\author{Calin-Andrei Pantis-Simut}
\thanks{These authors contributed equally to this work}
\affiliation{University of Bucharest, Faculty of Physics, Atomistilor 405, Magurele-Ilfov, 077125, Romania}
\affiliation{Research Institute of the University of Bucharest (ICUB), Mihail Kogalniceanu Blvd 36-46, Bucharest, 050107, Romania}
\affiliation{Horia Hulubei National Institute for Physics and Nuclear Engineering, Reactorului 30, Magurele-Ilfov, 077125, Romania}

\author{Amanda Teodora Preda}
\thanks{These authors contributed equally to this work}
\affiliation{University of Bucharest, Faculty of Physics, Atomistilor 405, Magurele-Ilfov, 077125, Romania}
\affiliation{Research Institute of the University of Bucharest (ICUB), Mihail Kogalniceanu Blvd 36-46, Bucharest, 050107, Romania}
\affiliation{Horia Hulubei National Institute for Physics and Nuclear Engineering, Reactorului 30, Magurele-Ilfov, 077125, Romania}

\author{Lucian Ion}
\affiliation{University of Bucharest, Faculty of Physics, Atomistilor 405, Magurele-Ilfov, 077125, Romania}

\author{Andrei Manolescu}
\affiliation{Department of Engineering, Reykjavik University, Menntavegur 1, IS-102 Reykjavik, Iceland}

\author{George Alexandru Nemnes} 
\email{Corresponding author: \\ G. A. Nemnes (nemnes@solid.fizica.unibuc.ro)}
\affiliation{University of Bucharest, Faculty of Physics, Atomistilor 405, Magurele-Ilfov, 077125, Romania}
\affiliation{Research Institute of the University of Bucharest (ICUB), Mihail Kogalniceanu Blvd 36-46, Bucharest, 050107, Romania}
\affiliation{Horia Hulubei National Institute for Physics and Nuclear Engineering, Reactorului 30, Magurele-Ilfov, 077125, Romania}

\begin{abstract}
	Accurate and efficient tools for calculating the ground state properties of interacting quantum systems are essential in the design of nanoelectronic devices. The exact diagonalization method fully accounts for the Coulomb interaction beyond mean field approximations and it is regarded as the gold-standard for few electron systems. However, by increasing the number of instances to be solved, the computational costs become prohibitive and new approaches based on machine learning techniques can provide a significant reduction in computational time and resources, maintaining a reasonable accuracy. Here, we employ {\tt pix2pix}, a general-purpose image-to-image translation method based on conditional generative adversarial network (cGAN), for predicting ground state densities from randomly generated confinement potentials. Other mappings were also investigated, like potentials to non-interacting densities and the translation from non-interacting to interacting densities. The architecture of the cGAN was optimized  with respect to the internal parameters of the generator and discriminator. Moreover, the inverse problem of finding the confinement potential given the interacting density can also be approached by the {\tt pix2pix} mapping, which is an important step in finding near-optimal solutions for confinement potentials. 
\end{abstract}

%
%
%
%
%


\maketitle

\section{Introduction}


Machine learning (ML) has found extensive applications in multiple research fields in the last decade, bringing along a new paradigm in science, based on a more efficient and versatile analysis of experimental and simulated data \cite{PUGLIESE202119,Alzubaidi2021}. Statistical models and high-end programming have led to the build-up of deep learning techniques that solve problems of clustering, regression and classification \cite{Sarker2021}. In particular, material science and nanotechnology have adapted ML algorithms in order to provide an accelerated interpretation of data and reduce the resources needed for materials \cite{Schmidt2019,doi:10.1021accountsmr.1c00238,Fu_2023} and device \cite{doi:10.1021/acsnano.8b04726,Kim2022} design, based on calculated examples or experimental results. The field of artificial intelligence has also extended to a variety of topics such as predicting molecular electron densities \cite{Rackers_2023}, reduction of the noise level in high-resolution electron microscopy images \cite{Gambini_2023} or more theoretical areas of condensed matter \cite{Bedolla_2021}, such as quantum phase transitions \cite{PhysRevB.97.134109} and learning topological invariants \cite{PhysRevLett.120.066401}.

The physics of nanoelectronic devices and quantum information applications relies heavily on an accurate and efficient description of many-body states. Traditionally, the many-body systems have been approached by mean-field theories like Hartree-Fock and density functional theory (DFT), the latter being mostly employed in the context of atomistic calculations. However, for applications that require q-bit level descriptions beyond mean-field approaches, computationally more demanding methods such as the exact diagonalization (ED) method \cite{PhysRevB.61.4718,PhysRevB.81.155442,PhysRevB.84.115311} are necessary. Many-electron states have been previously analyzed in quantum dot (QD) systems with top gate arrays \cite{Nemnes_2022}, where the exponential increase in the number of gate voltage configurations leads to a prohibitively large computation effort. Efficiently solving a large number of many-body Hamiltonian diagonalizations is typically required in the design of nanoelectronic devices and this is a suitable task for ML approaches.
	A lot of effort has been devoted to learning the electron densities, particularly in DFT frameworks, using the local atomic environment \cite{Kamal_2020}, equivariant graph neural networks \cite{Jorgensen2022} or by solving the many-body Schr\"odinger equation using a PauliNet as a deep-learning wavefunction Ansatz \cite{Hermann2020} or trial wave-functions implementing Pauli principle \cite{HAN2019108929}.

	Visualization has always been essential for the understanding and interpretation of the data. In the context of condensed matter, one idea is the use of graphs as means to encode the information about atomic and molecular structures \cite{PhysRevMaterials.4.093801,PhysRevLett.120.145301,SCHWARZER2019322}. Along with  the development of advanced deep learning methods, it also became possible to create algorithms that gain insights into raw representations such as pixels of an image. For this particular domain, convolutional neural networks (CNNs) have proved to be decisive. In material physics, CNNs have been employed for a variety of applications, from the prediction of the ionic conductivity of a ceramic material from image quality maps \cite{kondo2017} to the prediction of the space groups and the crystallographic dimensionality of thin film  materials from XRD spectral inputs \cite{Oviedo2019}  and a lot of work is now invested in explaining the mechanism through which CNNs make accurate predictions \cite{ml_rev_2022,cecen2018}.

	Autoencoders, which have CNNs embedded in their architecture, were used to learn low-dimensional representations of the data from a material database and subsequently incorporate it in a data-driven solver to improve efficiency \cite{HE2021114034}. Other methods, like flow-based models have achieved significant results for variational inference \cite{pmlr-v37-rezende15}, while high quality image synthesis was obtained using diffusion probabilistic models \cite{NEURIPS2020_4c5bcfec}. One step further from conventional convolutional networks, generative adversarial networks (GANs), are gaining importance lately. Several research articles have focused on using GANs for microstructure synthesis \cite{PhysRevE.101.043308, Hsu2021} and materials design, by capturing the characteristics of complex materials and learning the mapping between latent variables and the structure \cite{yang2018}.

Another remarkable network architecture that is suitable for image processing is the conditional generative adversarial network (cGAN), in which both the discriminator and generator are given additional information and, from this point of view, are trained in a conditional setting. cGANs have already been employed in the field of many-body physics to genenerate quantum state tomographies \cite{PhysRevLett.127.140502} and even simulate the dynamical correlators for many-body systems \cite{PhysRevResearch.4.033223}. They have also been proved to be efficient in the prediction of Ising spin configurations at temperatures outside the training data set \cite{Ising_cgan}.

	Based on this type networks, Isola \textit{et al} proposed an algorithm for image translation known as {\tt pix2pix}, which learns a loss function that adapts to the data and can be applied to a wide range of image processing related tasks. This type of model is a valuable tool in image dehazing tasks, which aim to for improve the quality of images and increase visibility \cite{Li2021}. {\tt Pix2pix}  is also already employed in the field of medical imaging to generate lesion images from tumor sketches for effective data augmentation\cite{Toda2022}, for the detection, colorization and classification of tumor images \cite{mehmood2022} and to generate synthetic CT images MRI radiotherapy planning \cite{TAHRI2022108}. In the field of industry research, {\tt pix2pix} was employed for the purpose of generating new images with surface quality defects, relevant in the production of metal workpieces \cite{HOLSCHER2022}. Due to the popularity of the algorithm, there is also considerable interest to increase the speed of training and improve its efficiency \cite{Lupion2022}.

In this paper, we investigate cGANs implemented in {\tt pix2pix} method for predicting many-body charge densities in the ground state, for randomly generated quantum systems. In the training process of the cGAN, the mapping is performed between the confinement potentials and the densities corresponding to Coulomb interacting systems, calculated by ED method. A similar mapping is performed to yield the non-interacting densities. In this way, the exact diagonalization is bypassed, which is a considerable advantage as diagonalizing many-body Hamiltonians becomes prohibitive when the number of systems grows too large. Using the {\tt pix2pix} mapping, an efficient and accurate prediction of the interacting densities is achieved for new test systems. In addition, we provide a proof-of-concept for the inverse problem, i.e. generating a potential from an input density. 

	The paper is structured as follows. In Section\ \ref{model}, the class of model systems is described. In the next section, the numerical implementations of the ED and {\tt pix2pix} methods are detailed and some measures for quality assessment of generated images are indicated. The results obtained for different types of pix2pix mappings, involving potentials, non-interacting and interacting densities, are discussed in Section\ \ref{results}. Subsequently, a discussion is provided in Section \ref{disc}, outlining the advantages and limitations of the current approach. The accuracies of the predicted densities are analyzed for several cGAN configurations and optimal configurations are identified. Moreover, the method is shown to produce accurate results for the inverse problem as well.

\section{Model systems}
\label{model}

\begin{figure}[t]%
\begin{center}
\includegraphics[width=0.5\linewidth]{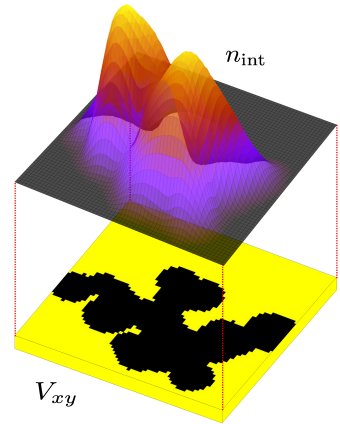}
\end{center}
	\caption{An interacting quantum system with random A-B type domains. The corresponding potential map in a typical configuration, $V_{xy}$, obtained using the procedure described in the text, is shown below. The system is defined on a two-dimensional square region of area $L \times L$, with vanishing boundary conditions for the wavefunctions. Starting from a step potential (yellow regions), $V_{\rm s}=0.5$ eV, a connected set of quantum wells (black regions), $V_0 = 0$ eV, defines the confinement potential for electrons.}
\label{system-description}
\end{figure}

The quantum systems consist of $N$ electrons confined in randomly generated potentials $V_{xy}$, defined on a two-dimensional square region, as depicted in Fig.\ \ref{system-description}. The potentials $V_{xy}$ correspond to connected groups of circular quantum wells (QWs) with different radii. These potential configurations resemble systems of interacting QDs such as two-dimensional self-assembled functionalized graphene QDs \cite{doi:10.1021/acsnano.1c09582}, randomly distributed QDs for memristive elements \cite{https://doi.org/10.1002/adma.201805284} or random geometric graphs of QDs \cite{nano11020375}. Monolayer graphene - hexagonal boron nitride films can form arbitrary shaped A-B type domains \cite{Levendorf2012,Geng2019}, where A and B are conductive and insulating domains, respectively. Moreover, the choice of random potential maps also ensures a thorough evaluation of the {\tt pix2pix} method.

	In order to get a balanced distribution of QWs in given potential map, the following scheme was considered. Starting with a flat potential step of height $V_{\rm s} = 0.5$ eV, a number of $N_{\rm qw} = 25$ flat QWs are placed inside the square region of linear size $L = 30$ nm, having the base potential $V_0 = 0$ eV. The centers of the QWs are randomly chosen, and their radii are uniformly distributed in the interval $L/16 < R < L/8$. When a new QW is added, it is allowed to partly overlap with the current QW, but no more than $3/4$ of its area. If the new QW is disconnected from the current QW, it is discarded and new values for the position and radius are chosen. The process continues until all $N_{\rm qw}$ are placed. Finally, if more than 80\% of the $L^2$ area is covered by the QWs ($V_0 = 0$ eV), the potential map is discarded and the process is started from the beginning. In this way, a connected ensemble of QWs is formed, with a high degree of variability. A total number of $N_{\rm V} = 5000$ potential instances are generated.

\section{Computational Methods}
\label{compmet}

\subsection{The exact diagonalization method}

The non-interacting one-body Hamiltonian for an electron in a two-dimensional confinement potential $V({\bf r})$ is:
\begin{equation}
	H_{0}=-\frac{\hbar^2}{2 m^{*}} \frac{\partial^2}{\partial \boldsymbol{r}^2}+V\left(\boldsymbol{r}\right),
\end{equation}
where $m^{*}$ is the effective mass and ${\bf r}\equiv(x,y)$ is the position vector in two-dimensions. 

The $N$-particle Hamiltonian is written as a sum of the single particle operators and the two-body operator, which describes the Coulomb interaction:

\begin{equation}
	H=\sum_{i=1}^N H_i+\frac{1}{2} \sum_{i}\sum_{j} V_{i j}, 
\end{equation}
where

\begin{eqnarray}
	& &H_i=H_0\left(\mathbf{r}_i\right), \\
	& &V_{i j}=V_C\left(\mathbf{r}_i, \mathbf{r}_j\right)=\frac{e^2}{4 \pi \epsilon_0 \epsilon_{r}} \frac{1}{\left|\mathbf{r}_i-\mathbf{r}_j\right|}.
\end{eqnarray}

The Hamiltonian in the second quantization becomes:

\begin{equation}
	\mathcal{H}=\sum_a \epsilon_a c_a^{\dagger} c_a+\frac{1}{2} \sum_{a b c d} V_{a b c d} c_a^{\dagger} c_b^{\dagger} c_d c_c,
\end{equation}
where $\epsilon_a$ are the energies of the single-particle states and the $V_{a b c d}$ coefficients corresponding to the Coulomb interaction are calculated based on the orbital components of the single-particle states, $\{\phi_{a, \sigma_z}\}$:
\begin{equation}
\begin{split}
	V_{a b c d}= \int d \mathbf{r} \int d \mathbf{r}^{\prime} \sum_{\sigma_z, \sigma_z^{\prime}} \phi_{a, \sigma_z}^*(\mathbf{r}) \phi_{b, \sigma_z^{\prime}}^*\left(\mathbf{r}^{\prime}\right) \\ 
\times \frac{ e^2}{4 \pi \epsilon_0 \epsilon_{r}} \frac{1}{\left|\mathbf{r}-\mathbf{r}^{\prime}\right|} \phi_{c, \sigma_z}(\mathbf{r}) \phi_{d, \sigma_z^{\prime}}\left(\mathbf{r}^{\prime}\right) .
\end{split}
\end{equation}

Solving the time independent Schr\"oedinger equation
\begin{equation}
\mathcal{H} \Psi_n=E_n \Psi_n,
\end{equation}
one obtains the eigenvalues $E_n$ and eigenvectors $\Psi_n \equiv \Psi_n\left(\mathbf{r}_1, s_1, \ldots, \mathbf{r}_{\mathrm{N}}, s_N\right)$. Then, the particle density in the ground state is:
\begin{eqnarray}
	& &n_0(\mathbf{r}) = N \sum_{s_1} \cdots \sum_{s_N} \int d\mathbf{r}_2 \cdots \nonumber\\
	& & \times \int d\mathbf{r}_N |\Psi_0\left(\mathbf{r}, s_1, \mathbf{r}_{\mathrm{2}}, s_2, \ldots, \mathbf{r}_{\mathrm{N}}, s_N\right)|^2 .
\end{eqnarray}

The numerical implementation of the ED method is described in detail in Ref.\ \cite{Nemnes_2022}. Choosing an appropriate single-particle basis, which fulfills the boundary conditions, we first solve the one-particle problem for a given two-dimensional potential, using a basis size $N_{\rm b}^2 = 32^2$, on a grid $N_{\rm x} \times N_{\rm y} = 64  \times 64$. Next, using the single-particle eigenfunctions, $\{\Phi_i(\mathbf{r})\}$, an $N$-particle basis of Slater determinants is assembled in the occupation number representation. By diagonalizing the two-particle Hamiltonian one obtains the ground state particle density:
\begin{equation}
	n_0({\bf r}) = \sum_k |C_{0k}|^2 \sum_{p=1}^{N} \left[ |\phi_{i_p(k),\uparrow}({\bf r})|^2 + |\phi_{i_p(k),\downarrow}({\bf r})|^2 \right],
\label{charge}
\end{equation}
where $C_{0k}$ is the expansion coefficient corresponding to the $k$-th Slater determinant and $\phi_{i_p(k),\uparrow}({\bf r})$, $\phi_{i_p(k),\downarrow}({\bf r})$ are the orbital components of the single-particle states $\Phi_{i_p(k)}(\mathbf{r})$, with spin up and spin down, respectively.

\subsection{cGAN implementation with {\tt pix2pix}}

\begin{figure*}[t]%
\begin{center}
\includegraphics[width=0.95\linewidth]{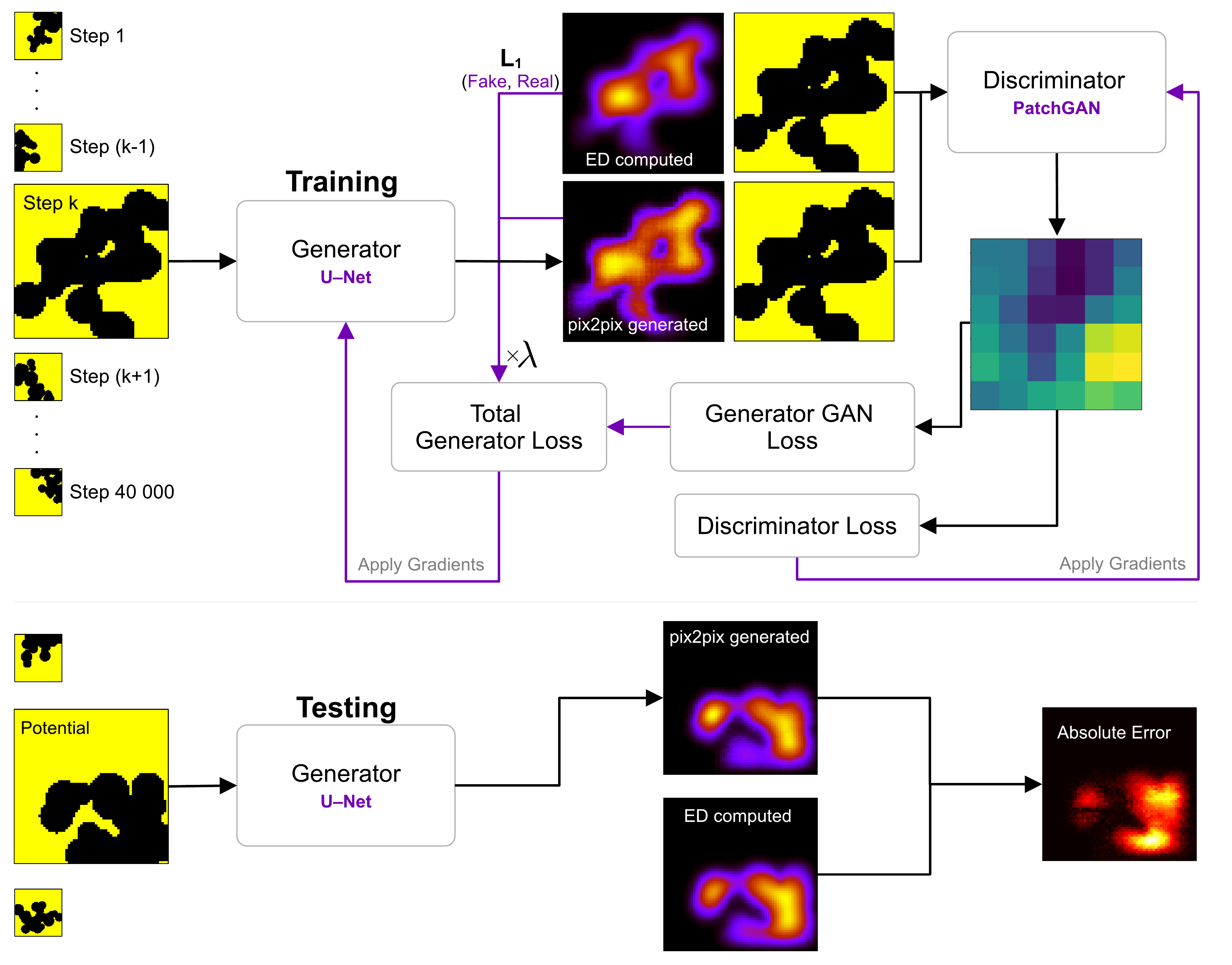}
\end{center}
	\caption{The workflow of the cGAN method, outlining the roles of the generator and discriminator networks, as a combined model, trained on (potential, density) pairs. The learning procedure for the total generator loss is indicated, according to the min-max problem described by Eq.\ \ref{Gtotloss}. In the testing phase, the generator network provides the predicted densities from the input potentials and these are compared with the reference densities obtained by ED method.}
\label{diagram}
\end{figure*}

The method developed by {\it Isola et al.} \cite{isola2016} makes use of a cGAN for general-purpose image-to-image translation. Like in other cGAN approaches, the generator-discriminator architecture of {\tt pix2pix} is set to optimize a global goal, namely that the generated output is made indistinguishable from the reference (ground truth). However, the main differences compared to other cGAN type approaches consists in the use of a U-Net architecture for the generator and a PatchGAN for the discriminator, which is more sensitive to local details. In Fig.\ \ref{diagram}, a diagram showing the training and testing phases of the cGAN is presented. 

We employ this approach to create three mappings of type $\alpha \mapsto \beta$, as follows:  ({\it i}) $V_{xy} \mapsto \tilde{n}_0$, ({\it ii}) $V_{xy} \mapsto \tilde{n}_{\rm int}$, ({\it iii}) $n_0 \mapsto \tilde{n}_{\rm int}$, where $V_{xy} = V(x,y)$ is the confinement potential, $n_0(x,y)$ and $\tilde{n}_0(x,y)$ are the calculated and generated non-interacting particle densities, respectively, and $\tilde{n}_{\rm int}(x,y)$ is the generated interacting particle density, which shall be compared to the calculated interacting particle density, $n_{\rm int}(x,y)$. 

Following the standard {\tt pix2pix} approach \cite{isola2016}, the generator $G$ performs a mapping from an input quantity-image  $\alpha \sim V_{xy}$ or $n_0$ to an output quantity-image $\beta \sim \tilde{n}_0$ or $\tilde{n}_{\rm int}$, except, of course, the trivial $n_0 \mapsto \tilde{n}_0$ mapping. The generator is trained to produce better and better images. On the other hand, the discriminator, $D$, is adversarially trained to classify the input it receives as {\it real} or {\it fake}. Previous cGAN methods \cite{10.1007/978-3-319-46493-0_20} employ a random (Gaussian) noise vector, denoted by $\gamma$, so we may describe the generator mapping as $G : \{\alpha,\gamma\} \mapsto \beta$. However, as the generator typically learns to ignore the noise introduced by the random vector $\gamma$, it is implemented in form of dropout in some layers of the generator and its overall influence is rather small. Therefore, the stochasticity of the predicted images by {\tt pix2pix} becomes negligible and can be completely excluded if the dropout is removed.   

In the original paper of {\it Isola et al.} \cite{isola2016} the objective function of the cGAN is expressed using the binary cross entropy as:
\begin{eqnarray}
	{\mathcal L}_{\rm cGAN}(G,D) &=& \mathbb{E}_{\alpha,\beta}[\log D(\alpha,\beta)] \nonumber \\ 
	 &+& \mathbb{E}_{\alpha,\gamma}[\log(1-D(\alpha,G(\alpha,\gamma))] .	
\end{eqnarray}
The objective $G^*$ is found as G tries to minimize ${\mathcal L}_{\rm cGAN}$ and D tries to maximize it and, in addition, an ${\mathcal L}_{L_1}$ loss, representing the difference between generated and reference images, is included:    
\begin{equation}
\label{Gtotloss}
	G^* = \arg \underset{G} \min\; \underset{D} \max\; {\mathcal L}_{\rm cGAN}(G,D) + \lambda {\mathcal L}_{L_1}(G),
\end{equation}	
where ${\mathcal L}_{L_1}(G) = \mathbb{E}_{\alpha,\beta,\gamma}[\|\beta - G(\alpha,\gamma)\|_1] $.
The parameter $\lambda=100$ sets the relative importance between ${\mathcal L}_{\rm cGAN}$ and ${\mathcal L}_{L_1}$.

In the min-max GAN problem, the generator loss would correspond to 
$\mathbb{E}_{\alpha,\gamma}[\log(1-D(\alpha,G(\alpha,\gamma)))]$, which should be minimized. However, the generator loss tends to saturate due to the vanishing gradients. This poses a challenge in training the generator at early stages, which causes the discriminator to outperform the generator and the model cannot optimally train. Instead, a non-saturating generator loss \cite{NIPS2014_5ca3e9b1} can provide a significant improvement:
\begin{equation}
\label{L_G}	
{\mathcal L}_{\rm G}^{\mbox{\tiny GAN}} = - \mathbb{E}_{\alpha,\gamma}[\log({\mathcal S}(D(\alpha,G(\alpha,\gamma))))], 
\end{equation}
where ${\mathcal S}(x)$ is the sigmoid function.
Then, the total generator loss is ${\mathcal L}_{\rm G}^{\mbox{\tiny GAN}} + \lambda {\mathcal L}_{L_1}(G)$.
The discriminator loss is defined as usual \cite{NEURIPS2018_e46de7e1}:
\begin{equation}
\label{L_D}	
{\mathcal L}_{\rm D}^{\mbox{\tiny GAN}} = 
				- \mathbb{E}_{\alpha,\beta}[\log ({\mathcal S}(D(\alpha,\beta)))]
				- \mathbb{E}_{\alpha,\gamma}[\log(1-{\mathcal S}(D(\alpha,G(\alpha,\gamma))))] .
\end{equation}

The architecture of the cGAN is specified by a number of parameters corresponding to the generator and discriminator networks. We shall first assume a typical configuration, called {\it reference configuration}, which is then modified for further optimizations. The cGAN translates grid-quantities set on $N_x \times N_y = 64 \times 64$ pixels, which are potentials and charge densities. The generator has an encoder-decoder configuration with 6 downsampling convolutional layers and 6 upsampling deconvolutional layers, all with strides $S_G = 2$, which keeps the size of the output equal to the size of the input. The discriminator receives two pairs of images, (input image, reference image) and (input image, generated image), which should be classified as real and fake, respectively. Its architecture includes 5 convolutional layers with the strides-sequence $S_D = (2,2,2,1,1)$, which reduces the input to an output of $6 \times 6$. This corresponds to a convoluted response of patch classification in real or fake, the patch size being dependent on the discriminator's architecture. The resulting patch size is $N_{\rm pt} \times N_{\rm pt} = 70 \times 70$ is larger than the size of the image, in which case the cGAN is referred to as ImageGAN \cite{isola2016}. Decreasing the number of layers in the discriminator, the patch size decreases in the sequence $N_{\rm pt} = 34, 16, 7, 4, 1$, where the limiting case with $N_{\rm pt} = 1$ is termed PixelGAN. The kernel size for both G and D is $\kappa = 4$. A comprehensive list of model parameters, training and testing procedures is presented in Table\ \ref{modelparam}.

\subsection{Error and accuracy measures for generated densities}

In many applications of image-to-image translation like e.g. {\it maps} $\mapsto$ {\it aerial photographs} or the opposite, a {\it perceptual validation} is often employed \cite{isola2016}. Assessing the quality of the generated images or comparing them with target images is generally not an easy task. 

A quantitative approach often employed is the structural similarity index measure (SSIM), which combines luminance, contrast and structure components \cite{1284395}. However, for the current aim of mapping the charge densities in quantum systems, the stochasticity of the model is limited and a strict comparison based on $L_1$, $L_2$ and $L_\infty$ norms also becomes a suitable assessment with a transparent interpretation. The $L_1$ norm reflects the amount of displaced charge in generated vs. reference systems, $L_2$ is related to the root mean squared error and $L_\infty$ corresponds to a local maximum error in the evaluation of the charge densities.  

In order to evaluate the difference between the generated and reference grid-based quantities, denoted by $\beta$ and $\beta_{\rm ref}$, we consider the $L_1$, $L_2$ and $L_\infty$ norms as possible measures, the first two being scaled by the number of grid points (pixels):
\begin{eqnarray}
L_1 &=& \frac{1}{N_x \times N_y} \; \| \beta - \beta_{\rm ref} \|_1	\\
L_2 &=& \frac{1}{\sqrt{N_x \times N_y}} \; \| \beta - \beta_{\rm ref} \|_2	\\
L_\infty &=&  \| \beta - \beta_{\rm ref} \|_\infty
\end{eqnarray}
On the other hand, SSIM can provide further assessment on the structural differences between the generated and target densities. In addition, we calculate a mean SSIM (MSSIM) employing a uniformly weighted $8 \times 8$ square window. In the subsequent analysis, for the calculation of SSIM and MSSIM we use the typical parameters suggested in Ref.\ \cite{1284395}. 

The prediction accuracy in an ensemble of $N_{\rm sys}$ generated and reference pairs, $\{(\beta_i,\beta_{{\rm ref},i})\}$, can be described by the $R^2$ coefficient of determination, calculated from the residual sum of squares, ${\rm SS_{res}}$, and the total sums of squares, ${\rm SS_{tot}}$ :
\begin{equation}
R^2 = 1 - \frac{\rm SS_{res}}{\rm SS_{tot}},
\end{equation}	
with
\begin{eqnarray}
	& & {\rm SS_{res}} = \sum_{i=1}^{N_{\rm sys}} \| \beta_{{\rm ref},i} - \beta_i \|_2^2, \\
	& & {\rm SS_{tot}} = \sum_{i=1}^{N_{\rm sys}} \| \beta_{{\rm ref},i} - \bar{\beta}_{{\rm ref},i} \|_2^2,
\end{eqnarray}	
where $\bar{\beta}_{\rm ref} = \frac{1}{N_{\rm sys}} \sum_{i=1}^{N_{\rm sys}} \beta_{{\rm ref},i}$. In the vector space of the grid-based quantities $\{\beta_i\}$, we define $\beta_i\pm\beta_j$ as pixel-wise addition and subtraction, respectively.


\section{Results}
\label{results}

\begin{figure}[t]%
\begin{center}
\includegraphics[width=0.85\linewidth]{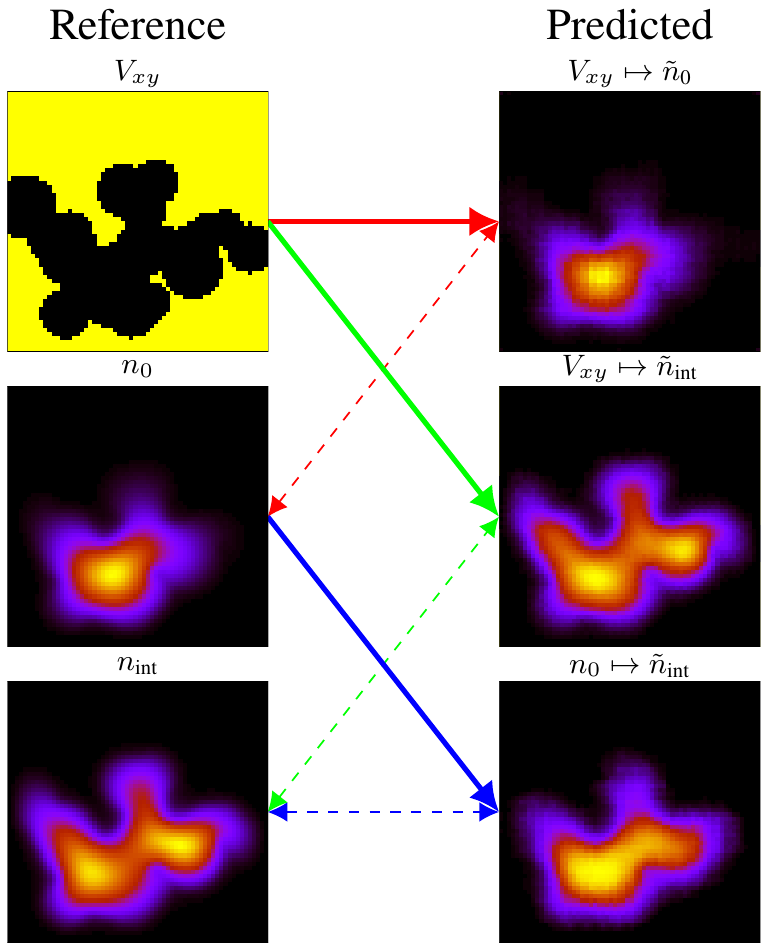}
\end{center}
	\caption{Non-interacting and interacting densities generated by the {\tt pix2pix} cGAN, for a two-particle system confined in a random potential $V_{xy}$: ({\it i}) $V_{xy} \mapsto \tilde{n}_0$, ({\it ii}) $V_{xy} \mapsto \tilde{n}_{\rm int}$, ({\it iii}) $n_0 \mapsto \tilde{n}_{\rm int}$ mappings, indicated by red, green and blue arrows, respectively. The left column shows the images of the calculated grid-based quantities: $V_{xy}$, $n_0$, $n_{\rm int}$. In the right column the generated images are depicted: $\tilde{n}_0$ and $\tilde{n}_{\rm int}$, the latter being determined from either $V_{xy}$ or $n_0$. The solid lines indicate the actual mapping, while the dashed lines indicate an association between the calculated (reference) data and generated (predicted) images.}
\label{3maps}
\end{figure}

\begin{figure}[t]%
\begin{center}
\includegraphics[width=1.0\linewidth]{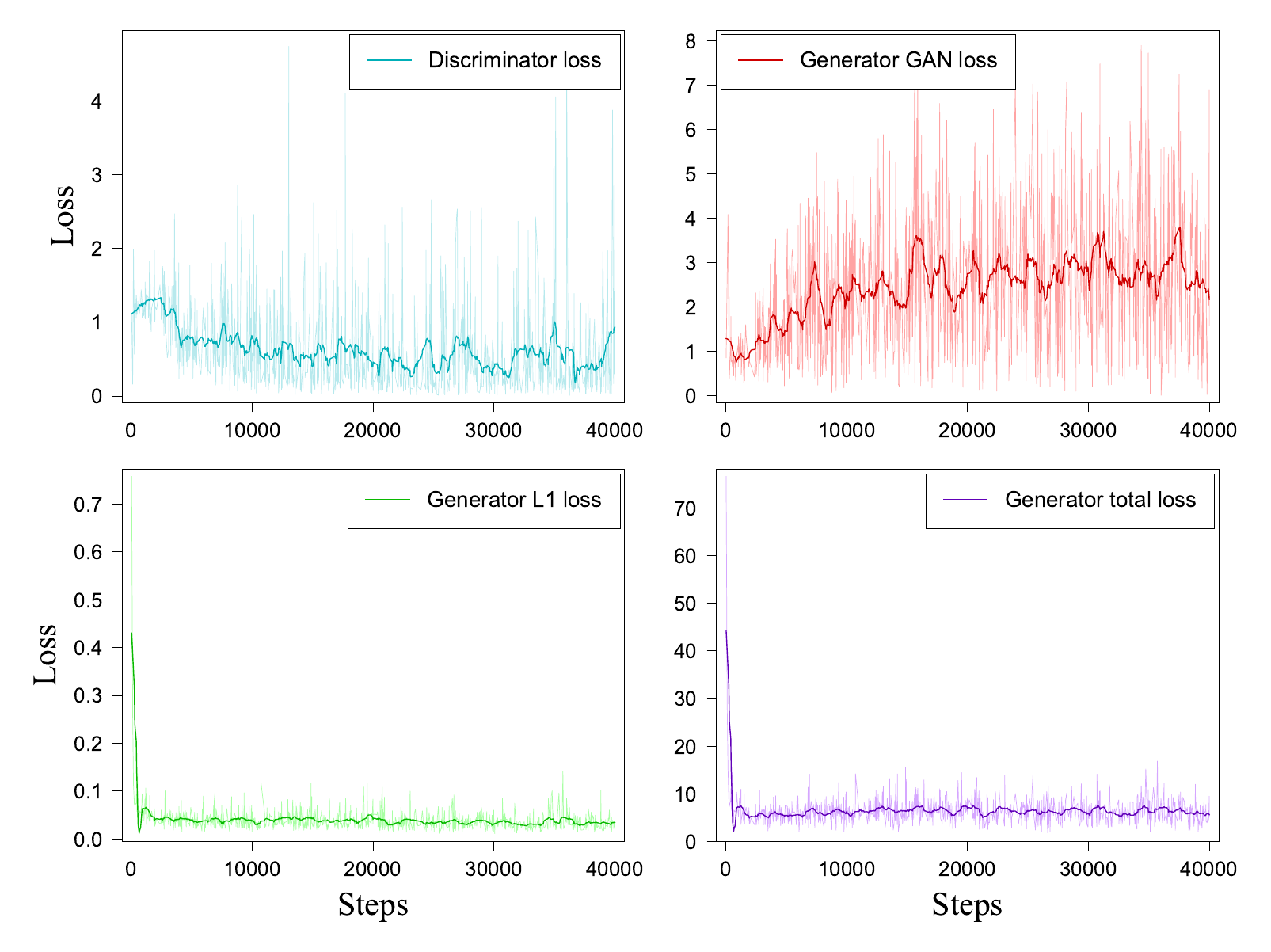}
\end{center}
	\caption{Generator and discriminator loss functions vs. the number of steps in a typical training run, for the mapping $V_{xy} \mapsto \tilde{n}_{\rm int}$ (thin lines). A smoothing is applied to all four data sets using the Savitzky-Golay filter with $3^{rd}$ degree polynomial and a window of 40 points, to better illustrate the trends (thick lines). In spite of the relatively large fluctuations, typical for cGAN architectures, the loss functions tend to stabilize.}
\label{lossGD}
\end{figure}

The quantum systems considered here consist of $N$ particles confined in randomly generated potentials $\{V_{xy}\}$ following the scheme described in Section\ \ref{model}. Starting with $N=2$ and using the reference cGAN configuration we perform the three mappings, as shown in Fig.\ \ref{3maps} for a typical instance: ({\it i}) $V_{xy} \mapsto \tilde{n}_0$, ({\it ii}) $V_{xy} \mapsto \tilde{n}_{\rm int}$, ({\it iii}) $n_0 \mapsto \tilde{n}_{\rm int}$, where the '$\sim$' symbol denotes generated quantities. The potentials $\{V_{xy}\}$ are readily available as input data, while the non-interacting densities, $\{n_0\}$, can be determined by one-particle calculations. The interacting densities, $\{n_{\rm int}\}$, are determined using the ED method, using the non-interacting many particle states obtained in the previous step, which are used to set up the two-particle basis. The first two mappings produce densities directly from the input potentials and, in particular, the second one, $V_{xy} \mapsto \tilde{n}_{\rm int}$, is of the highest importance, as it yields the interacting density without any diagonalization procedure after the model is trained. The third mapping starts from the non-interacting density, rather than the confinement potential, and it is performed for comparison.

Evaluated by visual inspection, all three mappings depicted in Fig.\ \ref{3maps} reproduce quite well the key features of the reference (calculated) densities. In a typical non-interacting calculation, the ground state charge density, $n_0$, is mostly localized in the quantum well region where the confinement is weaker, i.e. the wider part of the quantum well, so that the kinetic energy is minimized. In this case, the two electrons with opposite spins occupy the same space. However, when the Coulomb interaction is considered, the charge density in the ground state, $n_{\rm int}$, is more delocalized, being distributed in the quantum well of arbitrary shape, also in regions with stronger confinement. Qualitatively, the distribution of $n_{\rm int}$ is set by the tradeoff between the larger kinetic energy in stronger confinement regions and the Coulomb interaction between the particles occupying the same space in a region with weaker confinement. Fig.\ \ref{3maps} shows that the cGAN is able to learn the non-trivial features, so that the target quantities are reproduced with a high degree of accuracy. Additional examples are indicated in Fig.\ \ref{potential-mapping} and \ref{density-mapping} in the SI, for mappings from potentials and non-interacting densities, respectively.

\begin{figure}[h]%
\begin{center}
\includegraphics[width=0.99\linewidth]{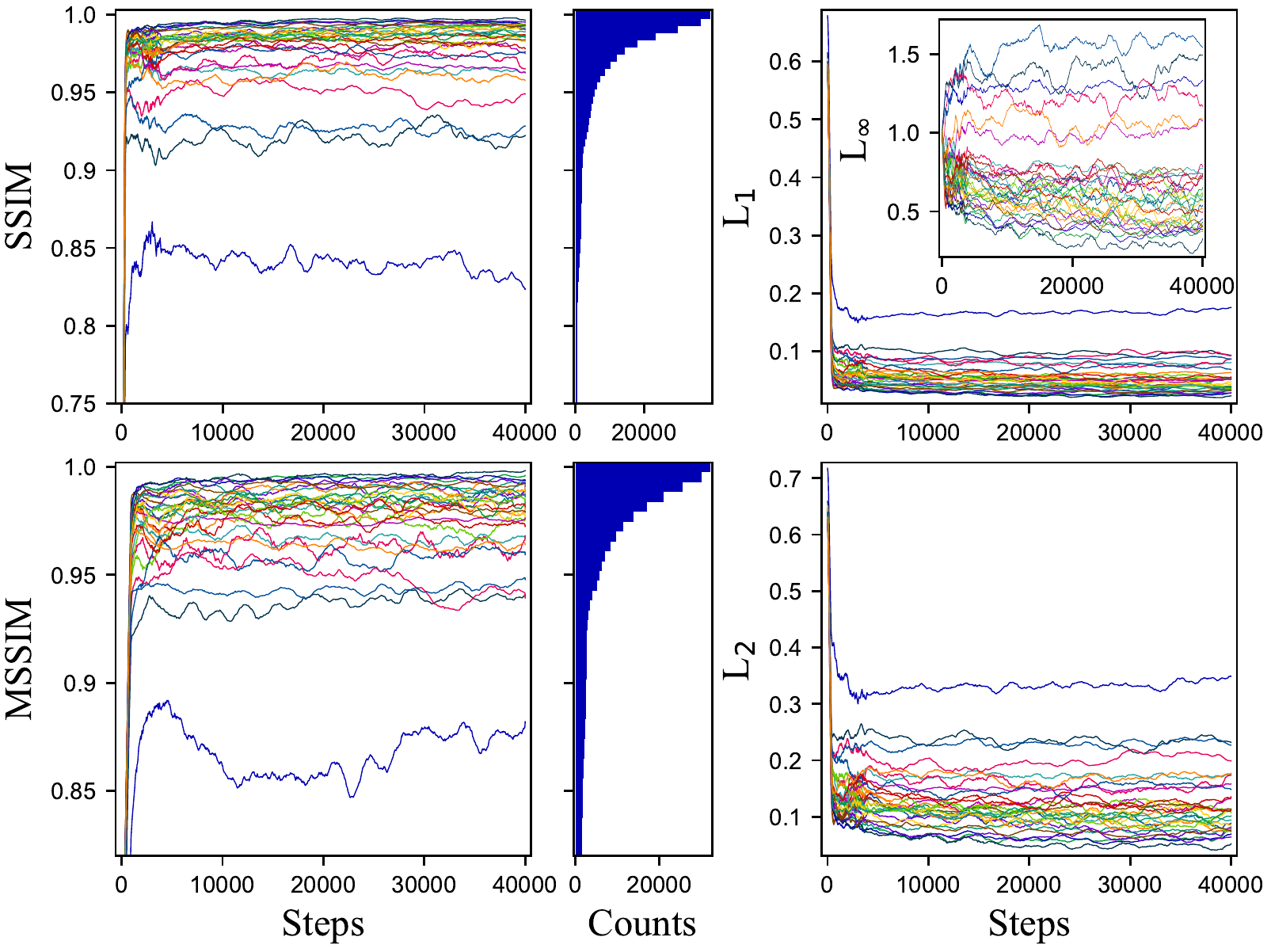}
\end{center}
	\caption{The evolution of SSIM and MSSIM during training for the mapping $V_{xy} \mapsto \tilde{n}_{\rm int}$, for a group of 30 test instances. Histograms of all SSIM and MSSIM values collected during training are depicted on the right hand sides of each plot. It is worth noting that the norms $L_1$, $L_2$ and, in part, $L_\infty$ are closely correlated with SSIM and MSSIM.}
\label{SSIM}
\end{figure}

\begin{figure}[h]%
\begin{center}
\includegraphics[width=0.99\linewidth]{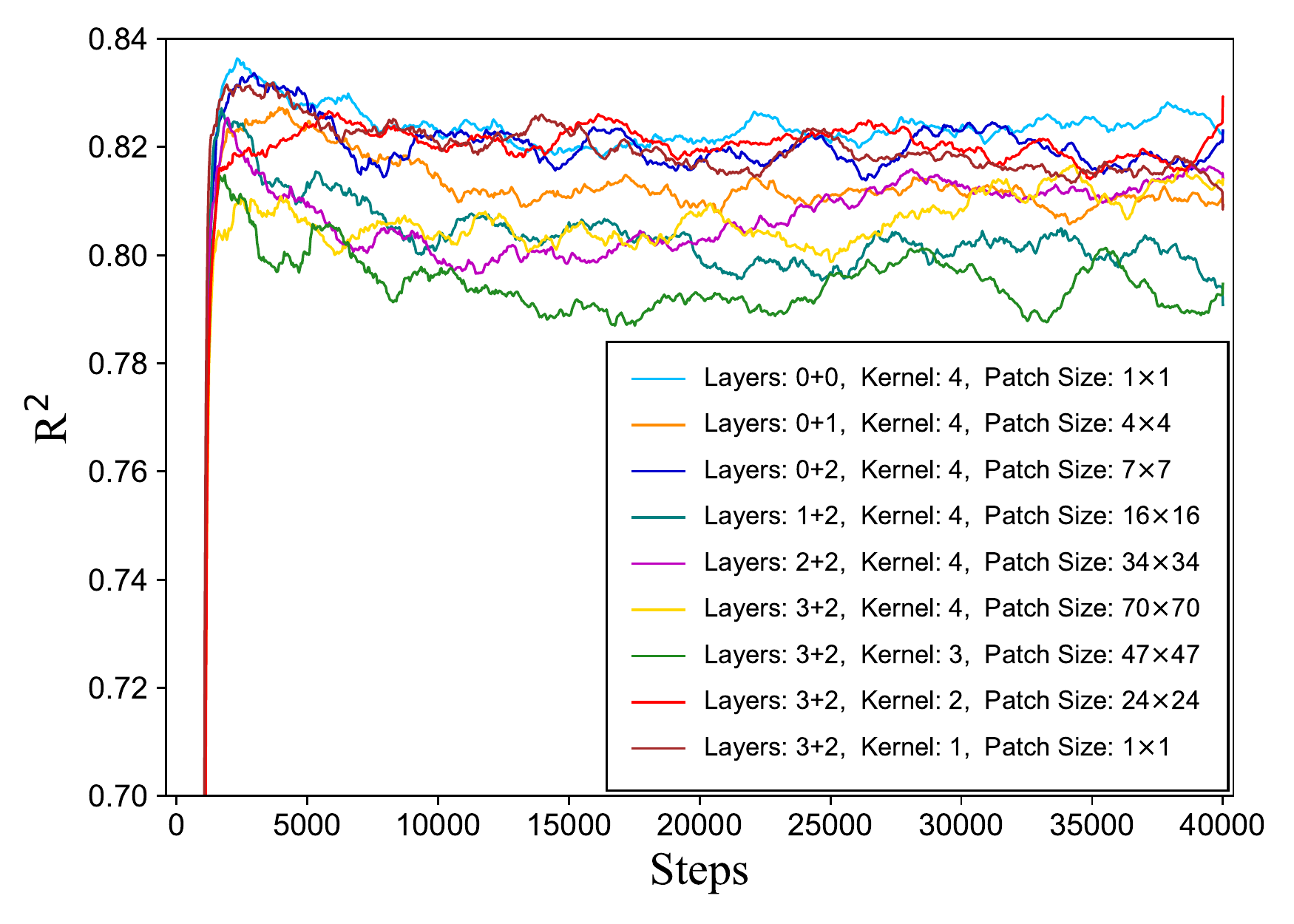}
\end{center}
	\caption{Prediction accuracies obtained with different cGAN architectures, based on tuning several key parameters, the number of convolutional layers in the discriminator network and the kernel size, resulting in the different patch sizes, $N_{\rm pt}$. Depending on $N_{\rm pt}$ values with respect to the image size, three groups of networks can be identified: ImageGAN ($N_{\rm pt}>64$), PatchGAN ($1<N_{\rm pt}<64$) and PixelGAN ($N_{\rm pt}=1$).}
\label{system}
\end{figure}

The training process of the cGAN was performed using $N_{\rm train} = 4800$ image pairs, using a batch size of 1 (instance normalization), while a number of $N_{\rm val} = 100$ and $N_{\rm test} = 100$ distinct samples were used for validation and test. For the relatively large training set, the averages of the potentials, non-interacting and interacting densities indicate a balanced distribution, as shown in Fig.\ \ref{mean-values} of the SI. These averages are later used to calculate the $R^2$ coefficient. 

During the training process, we monitor the loss functions of the generator and discriminator, which are depicted in Fig.\ \ref{lossGD} for a typical case. In contrast to the usual deep learning architectures, where the loss functions are specified, in cGANs the discriminator loss is learned from the input data, which usually brings large fluctuations. Therefore, instead of seeking the minima, the model becomes suitably trained when the loss functions are stabilized. This also poses a problem for the train-stopping-criterion, which is often optimized by visually checking a sequence of steps at the end of the training.

In order to assess the quality of the generated densities in the $V_{xy} \mapsto \tilde{n}_{\rm int}$ mapping, we calculate SSIM and MSSIM for a group of 30 samples from the test set and monitor their individual evolution as the model is trained. The data shown in Fig.\ \ref{SSIM} tends to overlap, indicating that high values (up to $\sim$ 0.9995) for both SSIM and MSSIM can be obtained, when the generated density becomes very similar to the target density, while at beginning of the training these values are below $\sim$0.3, when the first generated densities resemble the input potentials. However, a number of outliers are evidenced for which this procedure would produce somewhat worse results. These instances are described in Fig.\ \ref{outliers} of the SI. It is important to note that SSIM and MSSIM are in close correlation with the error measures based on $L_1$, $L_2$ and $L_\infty$, which are also represented in Fig.\ \ref{SSIM} for the same instances.

\begin{figure*}[t]%
\begin{center}
\includegraphics[width=0.85\linewidth]{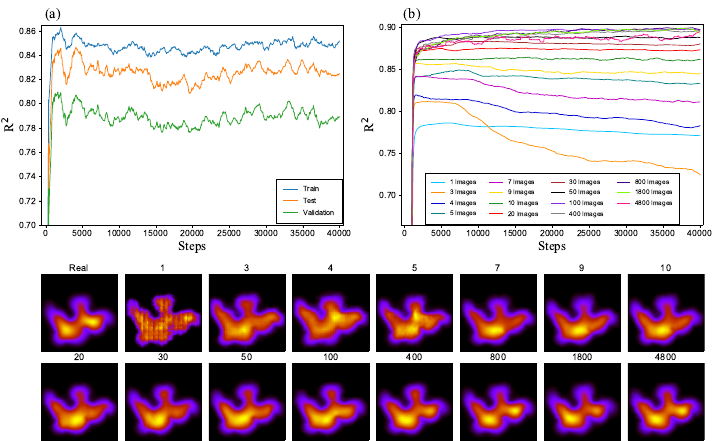}
\end{center}
	\caption{Accuracies measured by $R^2$ during training: (a) $R^2$ for training, validation and test sets, for the mapping $V_{xy} \mapsto \tilde{n}_{\rm int}$, with the standard cGAN configuration; (b) $R^2$ values for the test set, while varying the number of training examples, $N_{\rm train}$. The improvement of the final generated image for different sizes of the train sets is shown. For $N_{\rm train}<100$, the generated density merely resembles the potential (input image), while for $N_{\rm train}>800$ two individualized maxima can be observed, while further fine-tuning occurs for larger $N_{\rm train}$.}
\label{R2-tvt}
\end{figure*}

\begin{figure*}[t]%
\begin{center}
\includegraphics[width=0.85\linewidth]{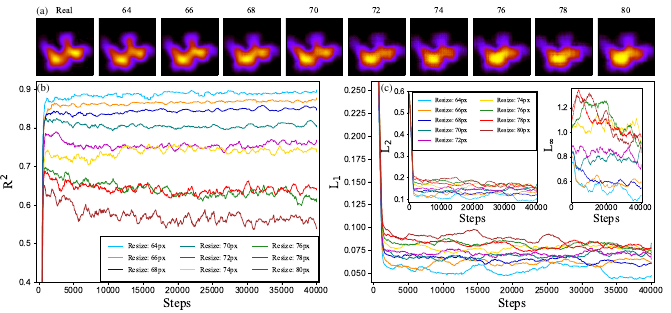}
\end{center}
	\caption{Analysis of the random jitter by applying resizing to $N_{\rm resize} \times N_{\rm resize}$ and then randomly cropping the images to the initial size. (a) The generated densities shows that the best results are obtained for no-resize (Resize = 64 px). (b) The $R^2$ coefficients and (c) the $L_1$, $L_2$, $L_\infty$ norms show consistently that the accuracy is reduced by increasing the resize parameter.}
\label{resize}
\end{figure*}

The overall accuracy of generated grid-based quantities on a set of examples is evaluated by the $R^2$ coefficient of determination. The evolution of $R^2$ for the test set vs. time step is depicted in Fig.\ \ref{system} for several cGAN architectures. We focus on the discriminator's architecture and vary the number of convolutional layers and the kernel size, which determines the patch sizes. The PixelGANs ($N_{\rm pt} =1$) perform better compared to an ImageGAN in the standard configuration, with 3+2 convolutional layers and a kernel $k=4$. However, overall, there are relatively small differences between all these configurations, with $R^2$ values in the interval 0.78 -- 0.84. 

Although, in contrast to standard (dense or convolutional) artificial neural networks, the utility of validation in GANs is questionable, we observe a systematic correlation between the training and a separate validation set, as indicated in Fig.\ \ref{R2-tvt}(a). This is particularly useful as one difficulty observed in the training of the cGANs consists in the sharp variations of the loss functions with the time step. The correlation between the training and validation sets enables us to optimize the training interval ($N_{\rm steps}$), i.e. it provides a stopping criterion so that the model produces accurate results. Then, the model is frozen and new densities are generated for the test set. Decreasing the number of input images the $R^2$ parameter is reduced, as one can see from Fig.\ \ref{R2-tvt}(b), while the relatively high values reflect the overall resemblance between the potential and the associated density.

\begin{figure}[t]%
\begin{center}
\includegraphics[width=0.85\linewidth]{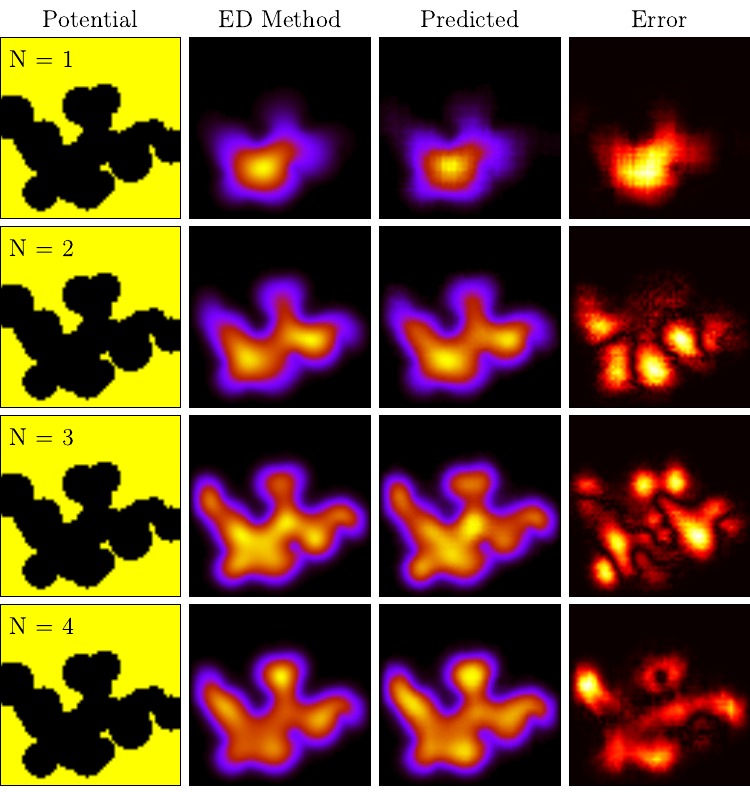}
\end{center}
	\caption{Mapping potentials to charge densities for different number of particles, $N=1,2,3$ and $4$, according to the mapping $V_{xy} \mapsto \tilde{n}_{\rm int}$. Increasing $N$, the charge densities have a larger spread towards the extremities of the confinement potential due to the Coulomb interaction, while several peaks develop. }
\label{largerNp}
\end{figure}

We also investigated possible optimizations of the cGAN approach. First, we account the effect of random jitter by resizing the images to $N_{\rm resize} \times N_{\rm resize}$ and then randomly cropping back to the original size, $64 \times 64$. This procedure was employed in a number of image translation problems discussed in Ref.\ \cite{isola2016}, like Map $\leftrightarrow$ aerial photograph, day $\rightarrow$ night images. In other cases, like black/white $\rightarrow$ color images no jittering was applied. A systematic investigation with respect to $N_{\rm resize}$ taking values from 64 to 80 in steps of 2 pixels shows that, for the $n_{\rm int} \mapsto \tilde{V}_{xy}$ mapping, no-resize ($N_{\rm resize}=64$) leads to the best results, $R^2 \sim 0.9$, and it decreases for larger $N_{\rm resize}$ values, as it can be seen in Fig.\ \ref{resize}. This is further confirmed by $L_1$, $L_2$ and $L_\infty$ norms, where the first two are well correlated, while, as expected, there are larger fluctuations for the $L_\infty$ norm. The quality of the generated images is also consistent with this trend, as the charge distribution becomes less diffuse. Secondly, we tested several values of the $\lambda$ parameter, which mixes ${\mathcal L}_{\rm G}^{\mbox{\tiny GAN}}$ and ${\mathcal L}_{L_1}$ losses and found a similar behavior as reported in Ref.\ \cite{isola2016}. Small values ($\lambda<50$) tend to produce image artifacts, e.g. misplaced peaks in the charge density, while large ones ($\lambda>100$) introduce broadening effects in the charge distribution. These trends are illustrated in Fig.\ \ref{lambda} in the SI.

Further, we employed this method for larger numbers of particles, namely $N=3$ and $N=4$. Fig.\ \ref{largerNp} shows the calculated and predicted ground densities for $N=1,2,3,4$, using the same potential as in Fig.\ \ref{3maps}. As the particle number increases, the number of many particle states also becomes considerably larger and the exact diagonalization calculations are even more computationally expensive. In the case of $N=3$, the number of many particle states is $N_{\rm MES} = 560$, while for $N=4$ we have $N_{\rm MES}=1820$. We did not impose any cut-off for the basis dimension in order to maintain the highest accuracy for the charge density calculation. The distribution of the training, validation and test data was maintained the same and the order of the randomly generated potentials was not altered. Also, the training parameters and the architecture of the networks remain unchanged. In this manner, we can compare the performance of the {\tt pix2pix} method as the number of electrons confined in the same potential configuration is varied. We focused only on the mapping $V_{xy} \mapsto \tilde{n}_{\rm int}$, predicting the interacting charge density from the two-dimensional potential. As more electrons are added to the two-dimensional system, we notice the number of maxima in the charge density profile increases, while the charge becomes more extended in the connected quantum wells. In the ground state, the expectation value of the Hamiltonian (the total energy) is minimized, so there is a competition in decreasing the Coulomb energy between the electrons and their kinetic energy, which is directly related to the effective confinement length. The interpretability of the many particle states is less obvious as one adds more fermions to the quantum system, particularly in the case of random potentials. However, the values of $R^2$ coefficients for three and four particles are 0.93 and 0.95, respectively. While the accuracy in identifying the charge density maxima slightly decreases as the number of particle is increased, the determination coefficient has higher values. This can be explained by the fact that the charge is more delocalized and it gradually takes the shape of the confining potential, which facilitates the prediction process. Additional instances illustrated in Fig.\ \ref{Np} further confirm these trends. For even larger number of particles one expects that the description of the quantum system becomes closer to a mean field approximation.

The inverse problem, i.e. mapping an input density to a generated potential, is highly important from both fundamental and technological perspectives. However, not every proposed ground state density can be obtained from a potential, which is known as the ${\mathcal V}$-representability problem \cite{doi:10.1073/pnas.76.12.6062}. Therefore, the inverse mapping $n_{\rm int} \mapsto \tilde{V}_{xy}$ is here performed starting from computed densities, rather than arbitrary ones. This provides a proof-of-concept for a solution to the inverse problem based on {\tt pix2pix} approach, if the target potential exists. As shown by Kohn in Ref.\ \cite{PhysRevLett.51.1596} small enough deviations from a ${\mathcal V}$-representable density is still in the same class, leading to a slightly different potential. 

\begin{figure}[h]%
\begin{center}
\includegraphics[width=0.75\linewidth]{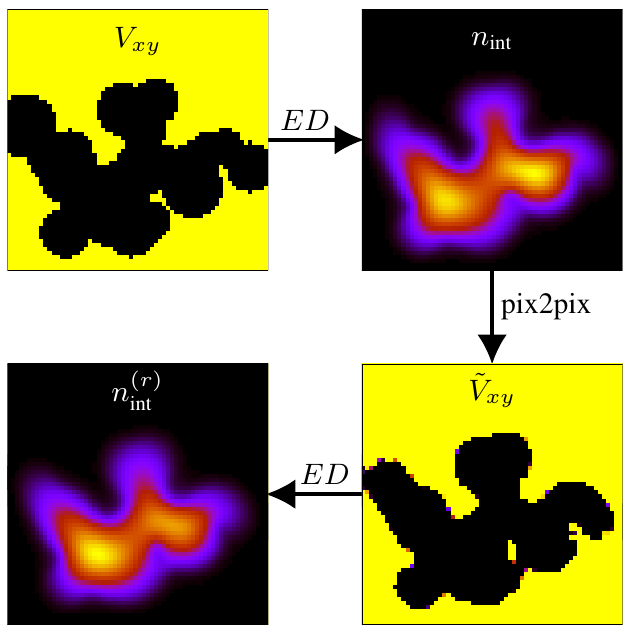}
\end{center}
	\caption{The inverse problem: generating potentials from interacting charge densities, according to the mapping $n_{\rm int} \mapsto \tilde{V}_{xy}$. Choosing an input potential, we calculate the interacting density by ED, which becomes the input image for the {\tt pix2pix} approach. The resulting potential, $\tilde{V}_{xy}$, is tested by computing its corresponding density, $n_{\rm int}^{(r)}$, which is very similar to the initial density $n_{\rm int}$, calculated from $V_{xy}$.}
\label{inverse-pb}
\end{figure}

A typical $n_{\rm int} \mapsto \tilde{V}_{xy}$, starting from an ED-computed density is shown in Fig.\ \ref{inverse-pb}. We use the same pair $(V_{xy},n_{\rm int})$, but this time $n_{\rm int}$ serves as input and the generated image contains the potential $\tilde{V}_{xy}$. Then, we recalculate the density corresponding to the generated potential, $\tilde{V}_{xy}$, which is denoted by $n_{\rm int}^{(r)}$. Comparing $\tilde{V}_{xy}$ with $V_{xy}$ and $n_{\rm int}^{(r)}$ with $n_{\rm int}$, i.e. generated vs. input quantities, one observes a large degree of similarity. To further support this, we plotted additional instances in Fig.\ \ref{density-mapping}(b) in the SI. There are still some small differences visible in the generated potentials compared to the original ones. In most cases, these differences occur for the regions with high confinement that are isolated from the main quantum well [e.g. as it is found in the instances 5 and 6 from Fig.\ \ref{density-mapping}(b) in the SI], which contain a small amount of localized charge. Consequently, as these QW regions are removed in the {\tt pix2pix}-generated potential by the cGAN model, the recalculated charge, $n_{\rm int}^{(r)}$, will not differ much from the input density, $n_{\rm int}$. Note that even the small islands present in some of the generated potentials are well represented compared to the originals. Then, as expected, the largest deviations occur at the boundaries, in particular at the edges of the square region, where the wavefunction vanishes. 

\begin{figure*}[t]%
\begin{center}
\includegraphics[width=0.95\linewidth]{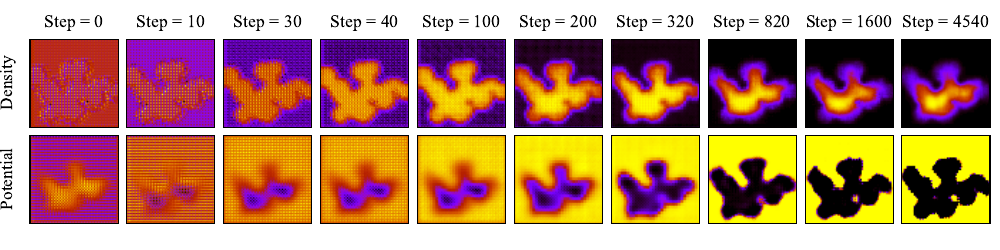}
\end{center}
	\caption{Evolution of the generated grid-based quantities $\tilde{n}_{\rm int}$ (first row) and $\tilde{V}_{xy}$ (second row), according to the mappings $V_{xy} \mapsto \tilde{n}_{\rm int}$ and $n_{\rm int} \mapsto \tilde{V}_{xy}$, respectively. In the two mappings, the initially generated images resemble the input potential and input density. Then, the images are gradually transformed, becoming more and more similar to the target density and potential, respectively.}
\label{evolution}
\end{figure*}

Although, in general, the ML methods are not very transparent with respect to their inner workings, it is interesting to observe the evolution of generated images representing densities and potentials. Fig.\ \ref{evolution} shows the sequential improvement of the generated images starting from the input images, as the model is improved. In the first row, the initial assumption for the density resembles the potential, with larger values outside the region corresponding to the quantum well. This is reversed in less than 10 steps and the charge is spread rather evenly inside the quantum well region. Starting with 200-300 steps, the density begins to localize inside the quantum well, while continuously changing its shape towards the target density, with two localized maxima. For the inverse problem, the evolution is shown in the second row of snapshots in Fig.\ \ref{evolution}. This time, the input is the interacting charge density and the first generated potential resembles it closely. However, in less than 10 steps, two quantum wells are individualized, then extending and merging in the first 100 steps. Subsequently, the shape of the generated potential becomes gradually closer to the target potential, which is depicted in Fig.\ \ref{inverse-pb}. The capacity of the method to reproduce the desired quantities is further confirmed by the SSIM values calculated for the pairs generated - reference, as shown in Fig.\ \ref{SSIM_n_Vxy} in the SI.

Overall, the {\tt pix2pix} approach provides an accurate and efficient alternative to predict the ground state density from the input potential or, conversely, to generate a potential from a given density, known to be ${\mathcal V}$-representable, once the cGAN is trained on a distinct set of calculated examples. Further investigations on excited states, as well as on quantum systems with larger numbers of particles can be pursued in a similar way. 

\section{Discussion}
\label{disc}

The proposed cGAN based on the {\tt pix2pix} method is an Ansatz-free approach, which makes no {\it a priori} assumption about the many-body wavefunction, which is usually done by setting Slater determinants or other trial functions. It directly provides a mapping between the random potentials and observable quantities like the ground state charge densities. Similarly, it could yield spin density maps for Hamiltonians which contain spin-orbit interaction and magnetic field contributions. The Ansatz-free approach may be advantageous particularly for problems where a suitable basis set is not easy to obtain, e.g. mesoscopic random structures. 

The mappings are achieved by learning a loss function that makes the generated images as close as possible to reality, which corresponds to the reference charge maps. This means that on a given class of potentials, the algorithm learns rather subtle features concerning the charge localization effects in the confinement potential, in the context of an interacting system. The generator loss combines the ${\mathcal L}_{\rm G}^{\mbox{\tiny GAN}}$ loss with the ${\mathcal L}_{L_1}$ loss. On one hand, the ${\mathcal L}_{\rm G}^{\mbox{\tiny GAN}}$ loss tends to produce rather sharp images resembling the reference ones, since these are not identified as fakes, but artifacts may be generated, e.g. with respect to the positioning of the charge density maxima. On the other hand, the ${\mathcal L}_{L_1}$ minimizes the average error per pixel with respect to the reference and typically results into more consistent and blurred images.

In its current form, the algorithm lacks stochasticity, which is an advantage for generating a deterministic output. Even though in other applications a highly stochastic generated output is desired, here the ground state charge density is uniquely determined by the confinement potential and particle number. In this respect, by removing the dropout in the generator layers, the {\tt pix2pix} method becomes fully deterministic. With respect to efficiency, similar to other ML approaches, the formulated problem should assume a relatively large set of many-body systems. A large variability in the class of potentials with respect to the total number of instances will pose a limitation to the current approach. Also, as the number of particles grows larger, the number of many particle states increases rapidly and the diagonalizations become feasible if an energy cut-off is imposed on the many-body basis set. 

Another important advantage of the current method is that the inverse problem, i.e. mapping the confinement potential from a given density, can be approached in a similar way. The ${\mathcal V}$-representability problem of the charge density can be circumvented by testing the generated potential. If the calculated density, using the ED technique, fits the input density well, then the generated potential can be adopted as a solution to the inverse problem, otherwise it shall be dismissed. Given the importance and difficulty of solving inverse quantum many-body problems the topic will certainly deserve future investigations.

\section{Conclusions}

We introduced an image-to-image translation approach based on the {\tt pix2pix} method to predict $N$-particle charge densities from the confinement potentials. The quantum systems are defined on two-dimensional square region with randomly generated potentials and the corresponding ground state densities are determined by exact diagonalization method. A large number of pair images is generated, corresponding to the confinement potentials and calculated interacting densities. Using the cGANs implemented in {\tt pix2pix} we perform three types of mappings: potential to non-interacting density, potential to interacting density and non-interacting to interacting density. Although all three mappings result in accurate predictions, the focus is on generating an interacting density from a given potential. Several cGAN architectures have been considered, by varying the number of convolutional layers and kernel size in the discriminator network. This analysis shows that a PixelGAN is most accurate, although other configurations yield comparable results. 

The possibility to perform an inverse mapping, i.e. starting from a density and generating a potential, is outlined. Here, we considered as input a calculated density, which ensures the ${\mathcal V}$-representability. The generated potential is then tested and confirmed by calculating the ground state density associated with it and comparing this density with the original one. 

The cGAN based approach provides an efficient solution for predicting non-interacting and interacting ground state densities when a large set of systems from a given class is required to be solved. Interestingly, the inverse problem can also be approached using this technique, which is important for the design of nanoelectronic devices. The {\tt pix2pix} method is shown to be accurate for describing interacting quantum systems and appears to be further well suited for a range of condensed matter problems.\\

\section*{Acknowledgments}
This work was supported by a grant of the Romanian Ministry of Research, Innovation and Digitalization, CNCS - UEFISCDI, project number PN-III-P4-ID-PCE-2020-1142, within PNCDI III.\\

\bibliographystyle{apsrev4-1}
\bibliography{manuscript_R1}


\appendix 
\renewcommand\thefigure{\thesection.\arabic{figure}} 
\setcounter{figure}{0}

\onecolumngrid


\section{Supplementary information}
\label{SI}
\setcounter{figure}{0}
\renewcommand{\textfraction}{0.05}  

 \begin{table}[!htbp] \centering
	 \caption{Architecture details and model parameters in the cGAN implementation (reference configuration) and prediction accuracy measures:}
 \label{modelparam}
\centerline{\begin{tabular}{@{\extracolsep{5pt}} cc} 
 \\[-1.8ex]\hline 
 \hline \\[-1.8ex] 
	   {\bf Model element / Method} & {\bf Description / Property / Value} \\
 \\[-1.8ex]\hline  \hline \\[-1.8ex] 
          1. Generator architecture&   \\
 \hline	 
	  Network type & Decoder-Encoder (U-Net) \\
	  Encoder & Number of layers = 6 \\ 
		  & Activation function = LeakyReLU, {\it slope} = 0.3 \\
		  & Input size = $64 \times 64 \times 1$  (one RGB channel)\\
		  & Convolution: {\it padding = same, strides = 2} \\
	  Decoder & Number of layers = 6 \\ 
		  & Activation function = ReLU ($\tanh$ for the last layer) \\
		  & Transposed Convolution: {\it padding = same, strides = 2} \\
		  & Dropout (optional): fraction = 0.5 \\
	  Generator (GAN) loss & ${\mathcal L}_{\rm G}^{\mbox{\tiny GAN}} = - \mathbb{E}_{\alpha,\gamma}[\log({\mathcal S}(D(\alpha,G(\alpha,\gamma))))]$,\\	
	          & for a more robust minimization of \\
		  & $\mathbb{E}_{\alpha,\gamma}[\log(1-{\mathcal S}(D(\alpha,G(\alpha,\gamma))))]$ \cite{NIPS2014_5ca3e9b1,NEURIPS2018_e46de7e1}\\
	  Total generator loss & ${\mathcal L}_{\rm G}^{\mbox{\tiny GAN}} + \lambda {\mathcal L}_{L_1}$,\\
	          & where ${\mathcal L}_{L_1}(G) = \mathbb{E}_{\alpha,\beta,\gamma}[\|\beta - G(\alpha,\gamma)\|_1] $ and $\lambda = 100$\\
 \\[-1.8ex]\hline  \hline \\[-1.8ex] 
          2. Discriminator architecture &   \\
 \hline	 
	  Network type & PatchGAN classifier \\
	               & Number of layers = 5 \\
		       & Input size = $64 \times 64 \times 2$  (potential-density pairs)\\
		       & Convolution: zero-padding, strides-sequence $S_D = (2,2,2,1,1)$ \\
		       & Patch size = $70 \times 70$ (ImageGAN)\\
	  Discriminator (GAN) loss & ${\mathcal L}_{\rm D}^{\mbox{\tiny GAN}} = 
				- \mathbb{E}_{\alpha,\beta}[\log ({\mathcal S}(D(\alpha,\beta)))]
				- \mathbb{E}_{\alpha,\gamma}[\log(1-{\mathcal S}(D(\alpha,G(\alpha,\gamma))))]$ \cite{NEURIPS2018_e46de7e1} \\	    
 \\[-1.8ex]\hline  \hline \\[-1.8ex]
          3. Training &   \\
 \hline	 
	  Optimizer & Adam optimizer, with {\it learning rate} = $10^{-4}$  \\
	            & and momentum parameters $\beta_1 = 0.5$, $\beta_2 = 0.999$ \\
	  Batch size & 	Instance normalization (batch size = 1) \\   
	  Initializer & Gaussian distribution, with zero mean\\
	              & and standard deviation 0.02 \\
 \\[-1.8ex]\hline  \hline \\[-1.8ex]
          4. Prediction accuracy measures&   \\
 \hline	 
 	  $R^2$ & Coefficient of determination for image data sets \\
	  SSIM & Structural Similarity Index Measure \cite{1284395} \\
	  MSSIM & mean-SSIM, using a window of $8 \times 8$ pixels \\ 
 \\[-1.8ex]\hline  \hline \\[-1.8ex]
\end{tabular}}
 \end{table}



\begin{figure}[h]%
\hfill
\includegraphics[width=0.48\linewidth]{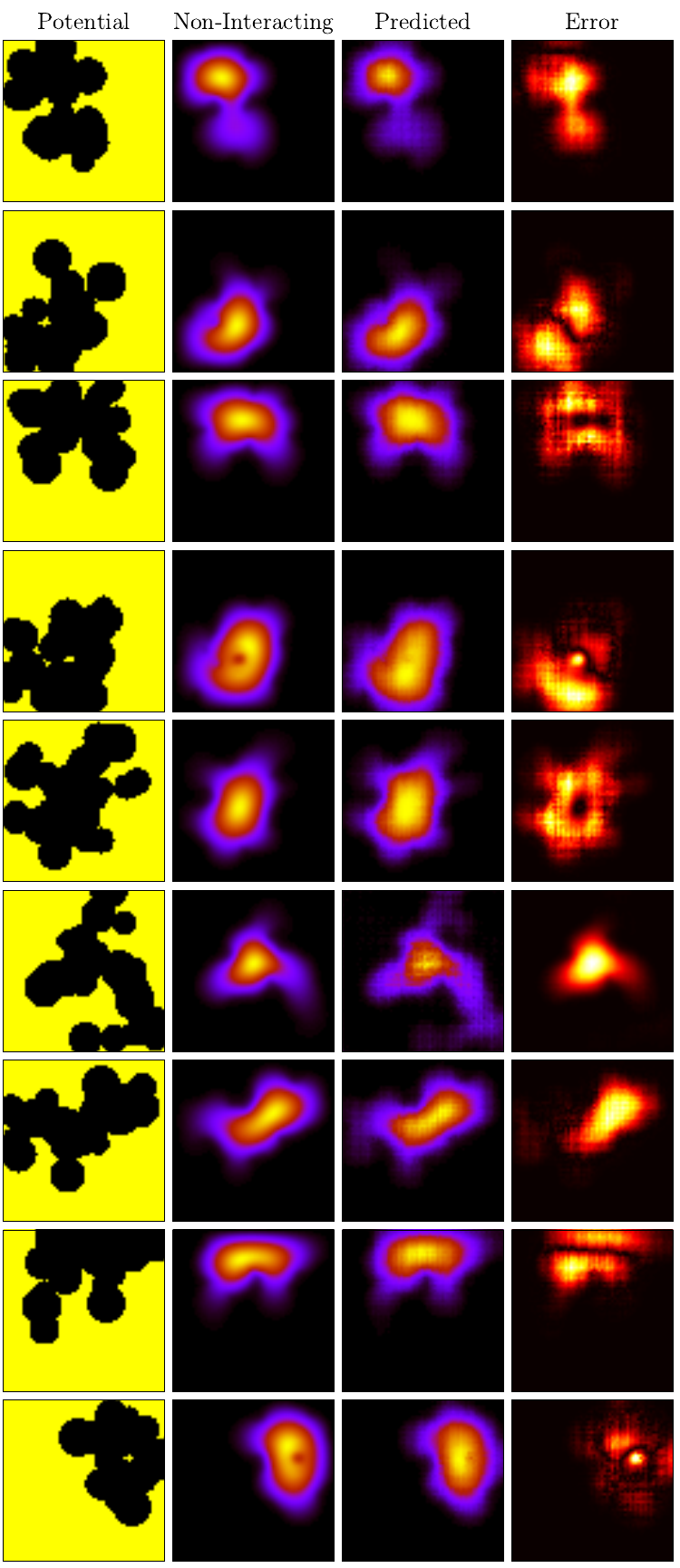}
\includegraphics[width=0.48\linewidth]{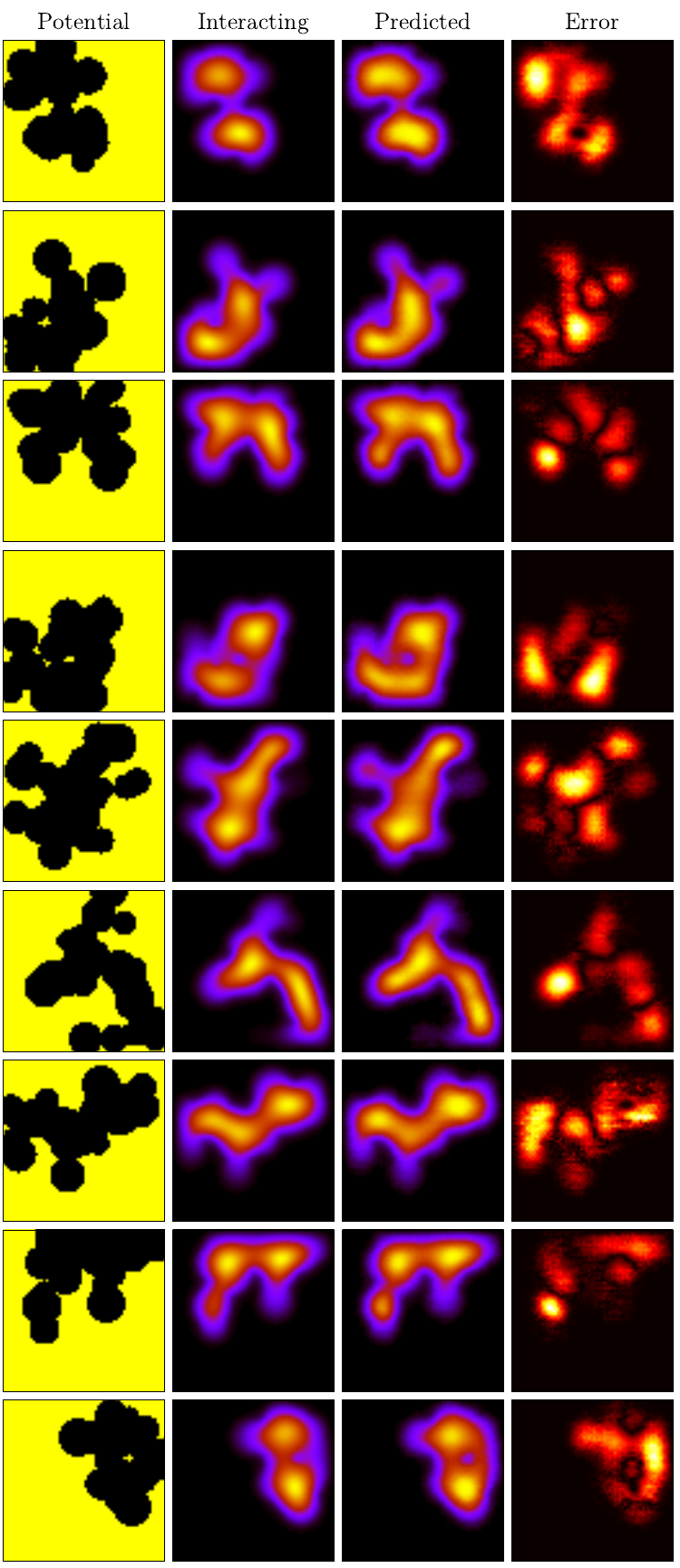}
\caption{ Mapping from potentials to (left) non-interacting and (right) interacting densities. The error maps correspond to differences between the target and predicted distributions (in absolute value).}
\label{potential-mapping}
\end{figure}

\begin{figure}[h]%
\hfill
\includegraphics[width=0.45\linewidth]{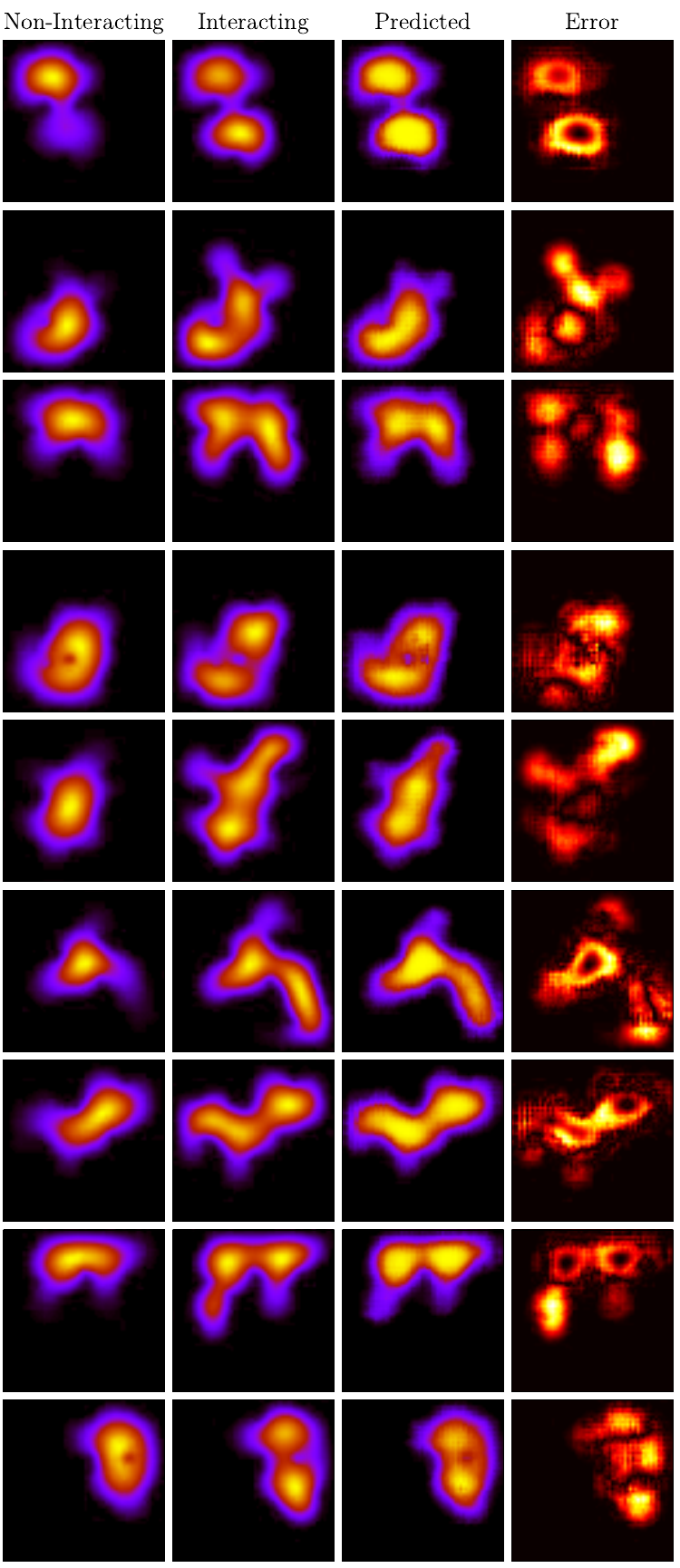}
\includegraphics[width=0.45\linewidth]{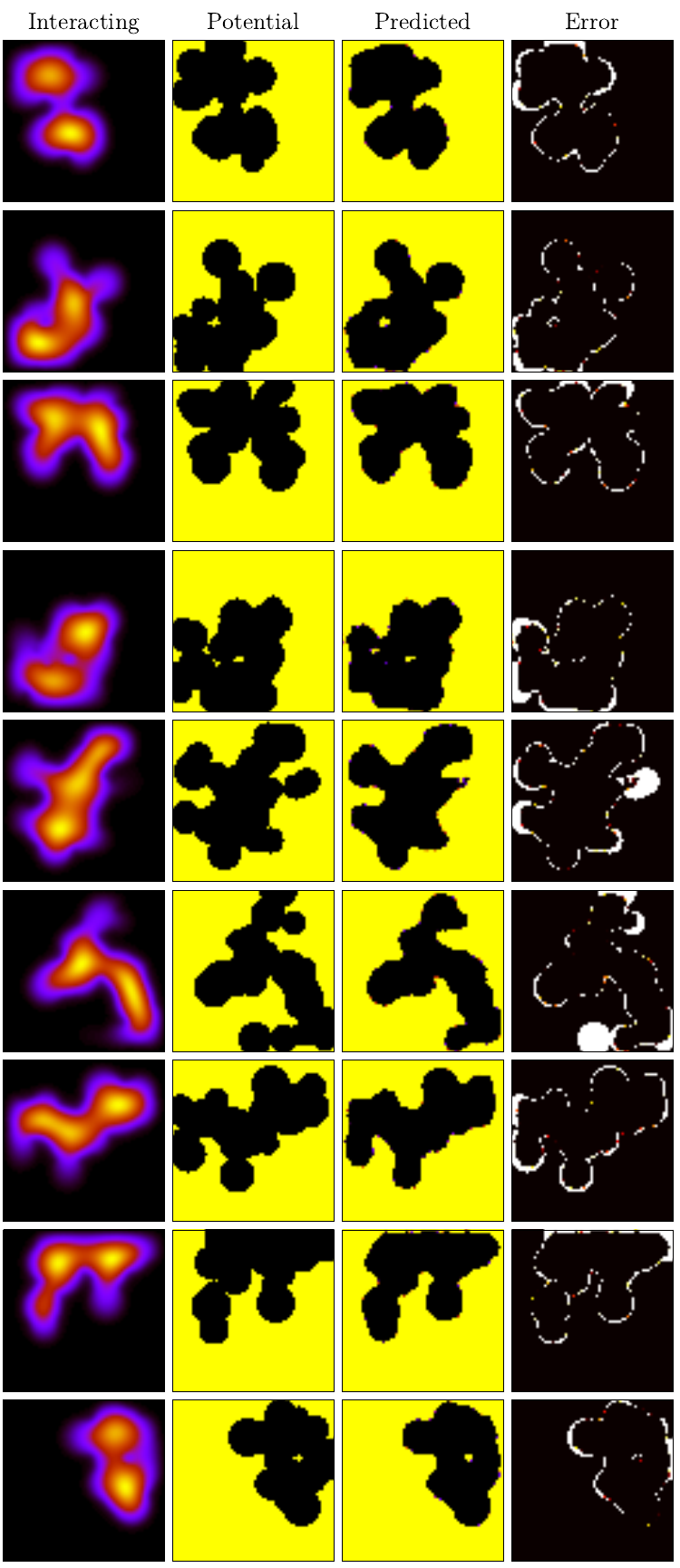}
\caption{ Mapping from charge densities: (left) non-interacting to interacting densities, $n_{0} \mapsto \tilde{n}_{\rm int}$ and (right) the inverse problem, $n_{\rm int} \mapsto \tilde{V}_{xy}$. The error maps correspond to differences between the target and predicted distributions (in absolute value).}
\label{density-mapping}
\end{figure}

\begin{figure}[t]%
\begin{center}
\includegraphics[width=0.75\linewidth]{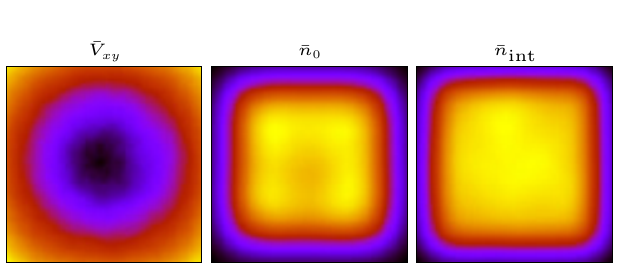}
\end{center}
	\caption{Averages of confinement potential ($\bar{V}_{xy}$), non-interacting density ($\bar{n}_0$) and interacting density ($\bar{n}_{\rm int}$) calculated using the training set ($N_{\rm train}=4800$). These average maps are used in the calculation of $R^2$. All three images indicate the balanced distribution of potential shapes. The average non-interacting density is more concentrated in the center of the square compared to the interacting density.}
\label{mean-values}
\end{figure}

\begin{figure}[t]%
\begin{center}
\includegraphics[width=0.75\linewidth]{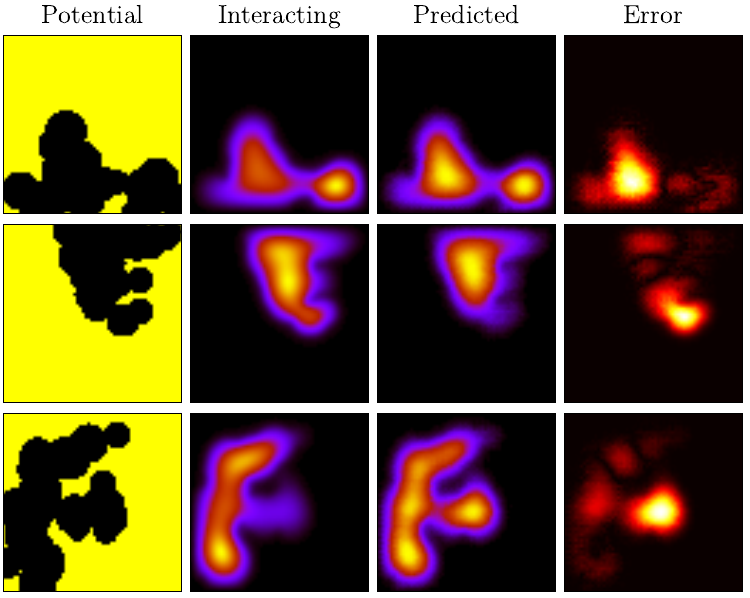}
\end{center}
	\caption{Three examples of outliers, which exhibit the largest deviations from the reference, as identified by the SSIM analysis in Fig.\ \ref{SSIM}.}
\label{outliers}
\end{figure}

\begin{figure}[h]%
\begin{center}
\includegraphics[width=0.45\linewidth]{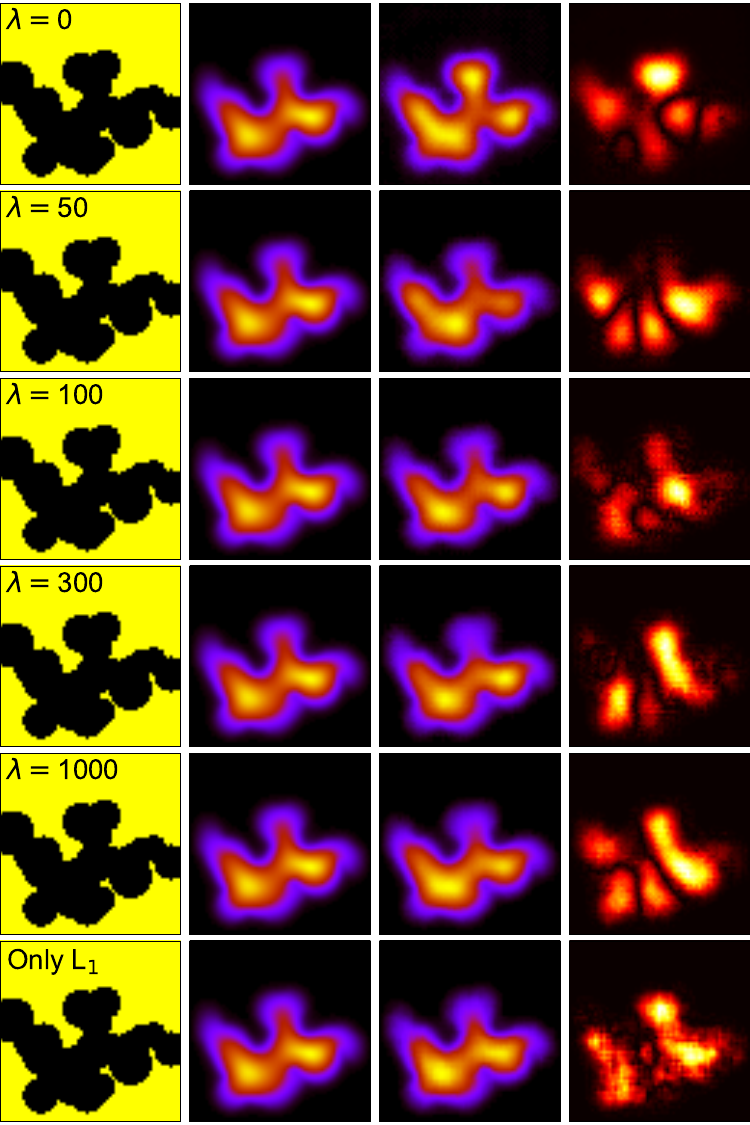}\hspace*{0.5cm}
	\includegraphics[width=0.45\linewidth]{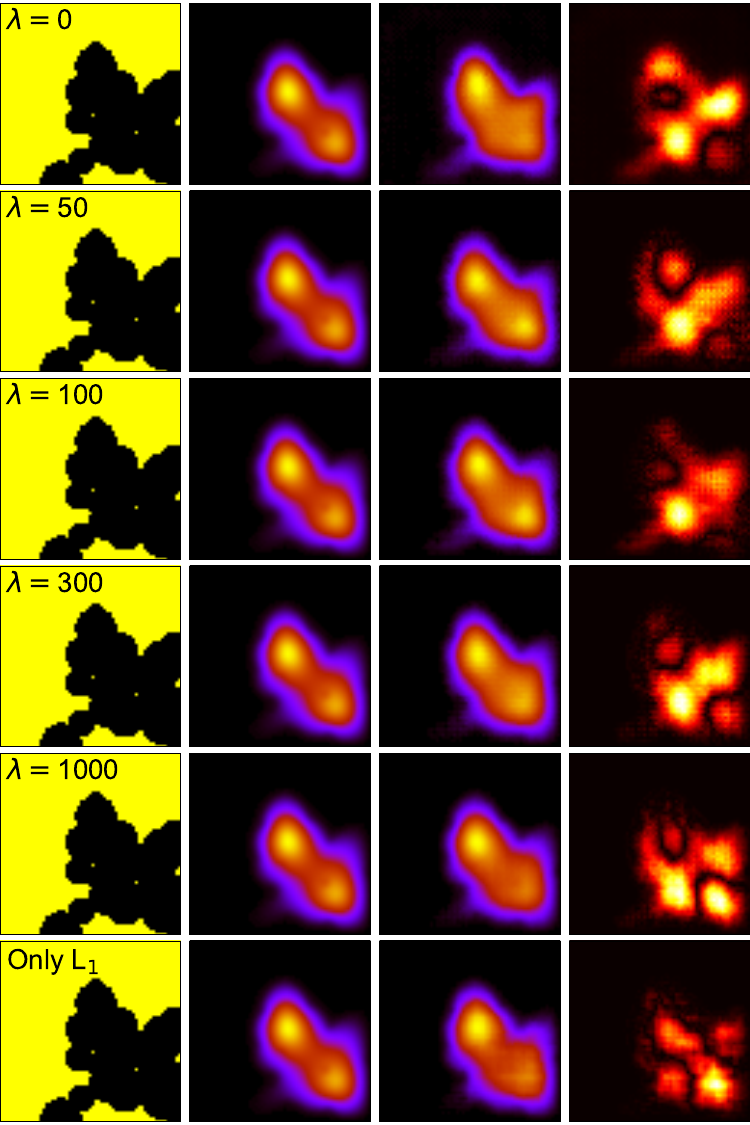} \\
\end{center}
	\caption{The effect of the ${\mathcal L}_{\rm G}^{\mbox{\tiny GAN}}$ and ${\mathcal L}_{L_1}$ mixing on the total generator loss. We considered $\lambda = 0, 50, 100, 300$ and $1000$, which includes the limiting cases of sole ${\mathcal L}_{\rm G}^{\mbox{\tiny GAN}}$ contribution ($\lambda = 0$) and sole ${\mathcal L}_{L_1}$ contribution. The optimal values are identified around $\lambda \approx 100$. Two different potential instances are presented. Smaller values tend to produce artifact-localizations of the charge density, while large values introduce an artificial broadening of the charge distributions.}
\label{lambda}
\end{figure}

\begin{figure}[h]%
\begin{center}
\includegraphics[width=0.45\linewidth]{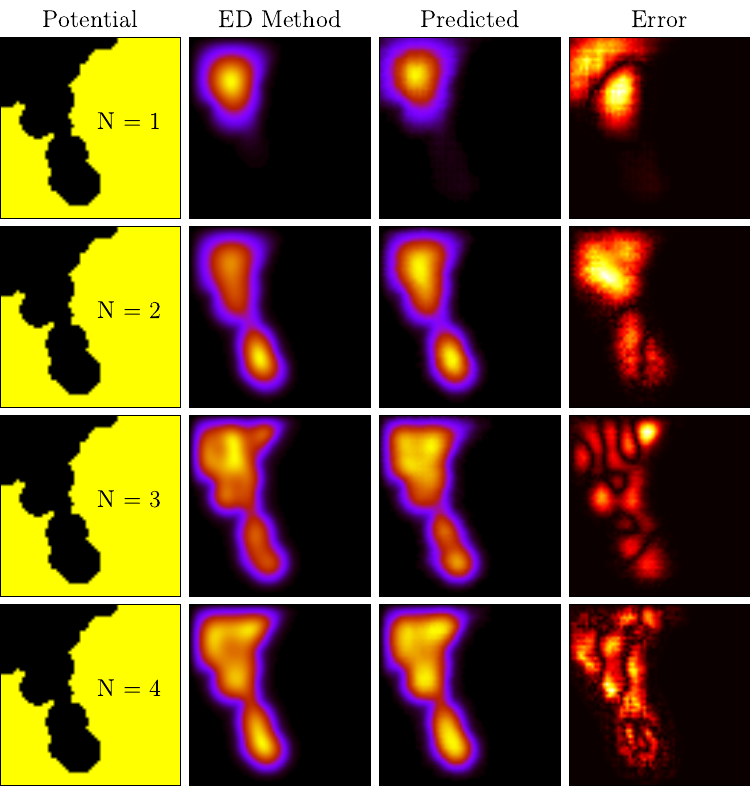}\hspace*{0.5cm}
	\includegraphics[width=0.45\linewidth]{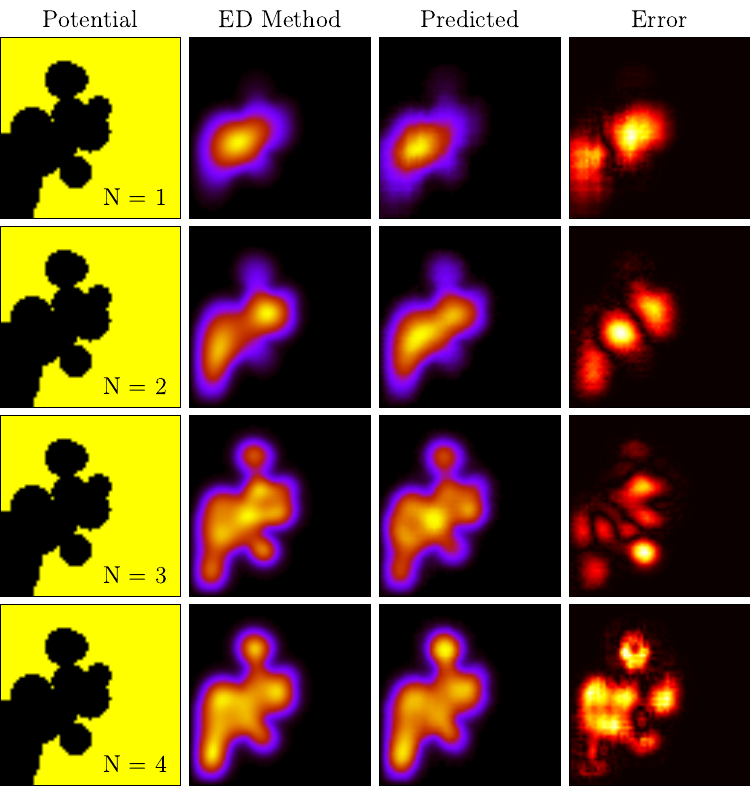} \vspace*{1cm}\\
\includegraphics[width=0.45\linewidth]{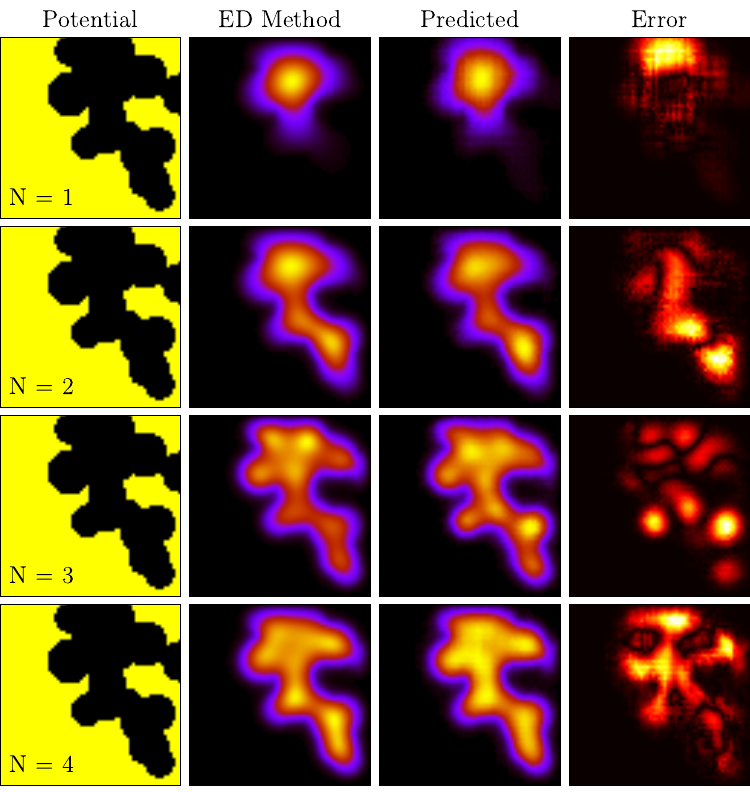}\hspace*{0.5cm}
\includegraphics[width=0.45\linewidth]{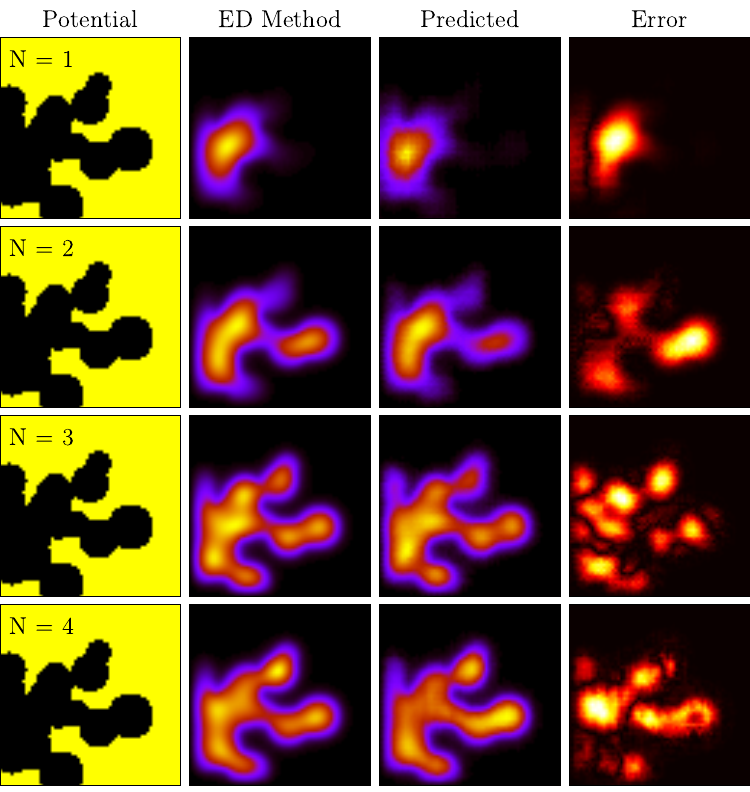} \\
\end{center}
	\caption{Mapping potentials to interacting charge densities for different number of particles, $N=1,2,3$ and $4$, considering four different potentials. }
\label{Np}
\end{figure}

\begin{figure}[h]%
\begin{center}
\includegraphics[width=0.45\linewidth]{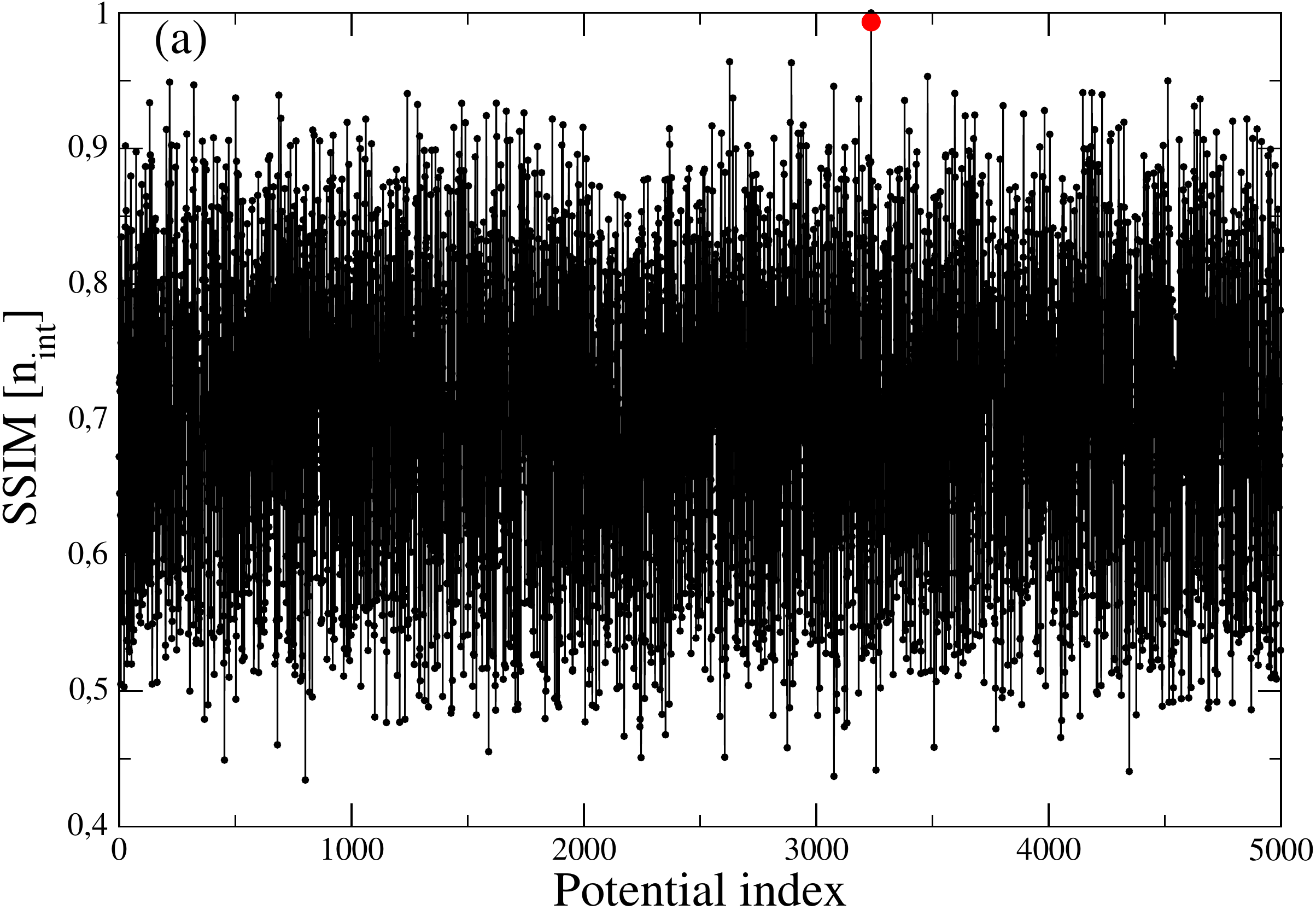}\hspace*{0.5cm}
\includegraphics[width=0.45\linewidth]{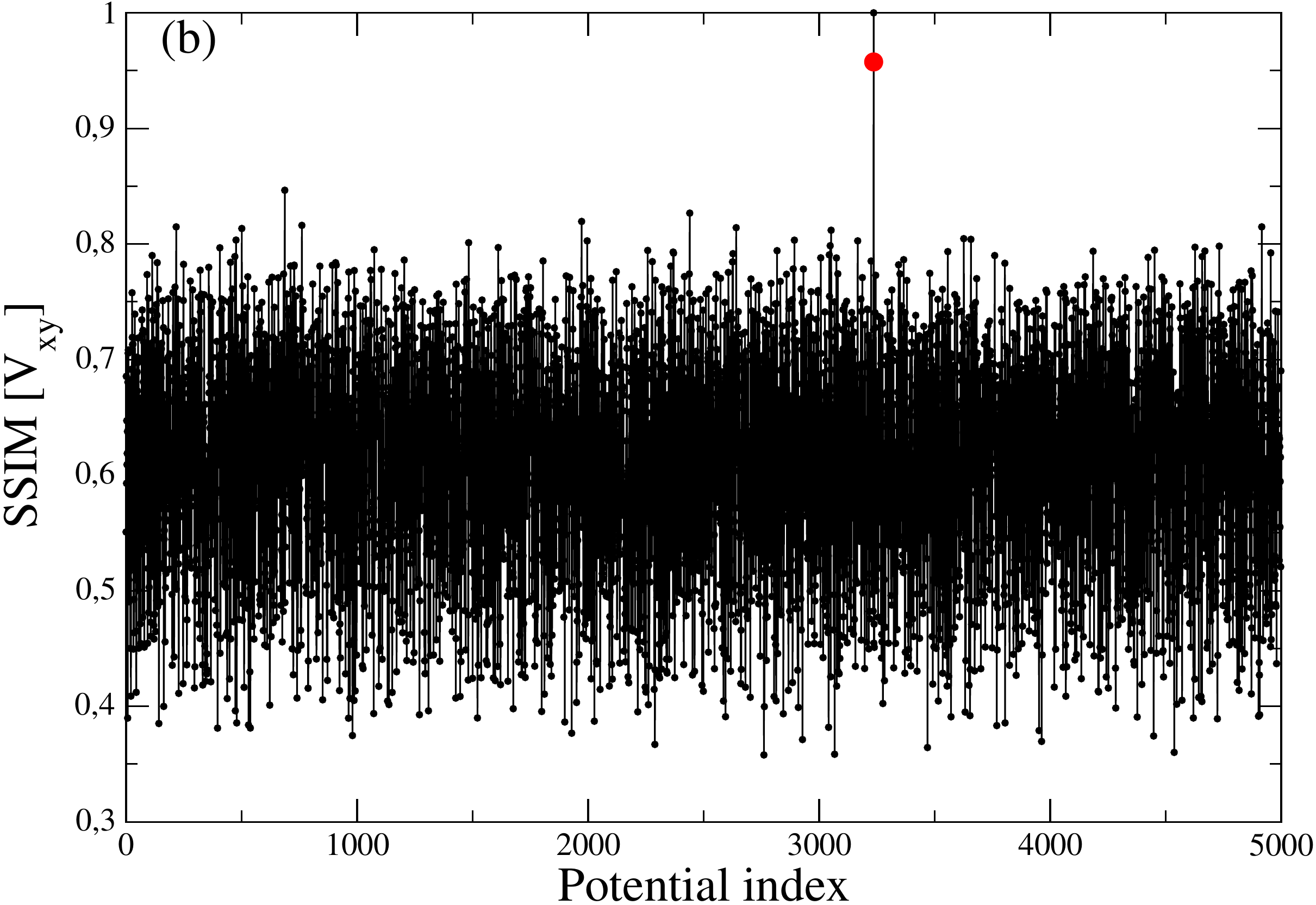}
\end{center}
	\caption{SSIM values for pairs of densities and potentials. One pair consists of the reference instance (index $i_0 = 3236$) from the test set, described in Figs.\ \ref{3maps} and \ref{inverse-pb}, and one other instance in the set of 5000 instances: (a) $(n_{{\rm int},i_0},n_{{\rm int},i})$ and (b) $(V_{xy,i_0},V_{xy,i})$, depicted by black dots. For $i=i_0$ we have SSIM = 1. The red dots indicate the comparisons between reference and generated quantities, for (a) densities $(n_{{\rm int},i_0},\tilde{n}_{{\rm int},i_0})$, SSIM=0.993 and (b) potentials $(V_{xy,i_0},\tilde{V}_{xy,i_0})$, SSIM=0.957, showing that the generated density ($\tilde{n}_{{\rm int},i_0}$) and potential ($\tilde{V}_{xy,i_0}$) have higher similarity with their references compared to any other instance in the set.}
\label{SSIM_n_Vxy}
\end{figure}

\end{document}